\documentclass[pdflatex,sn-mathphys-num]{sn-jnl}

\usepackage{graphicx}%
\usepackage{multirow}%
\usepackage{amsmath,amssymb,amsfonts}%
\usepackage{amsthm}%
\usepackage{mathrsfs}%
\usepackage[title]{appendix}%
\usepackage{xcolor}%
\usepackage{textcomp}%
\usepackage{manyfoot}%
\usepackage{booktabs}%
\usepackage{algorithm}%
\usepackage{algorithmicx}%
\usepackage{algpseudocode}%
\usepackage{listings}%
\usepackage{subcaption}
\usepackage{hyperref}
\usepackage{adjustbox}
\usepackage{caption}
\usepackage{bm}
\usepackage[hypcap=true]{caption}
\usepackage{minted}  

\usepackage[T1]{fontenc}            
\usepackage[utf8]{inputenc}         
\usepackage{microtype}              

\usepackage[most]{tcolorbox}        
\usepackage{fancyvrb}               

\definecolor{ExecutionPromptBack}{RGB}{255,250,210}
\definecolor{ExecutionPromptFrame}{RGB}{220,200,120}
\definecolor{ExecutionTitleBack}{RGB}{255,240,180}  
\definecolor{ExecutionPromptTitle}{RGB}{70,50,20}         
 
\newtcblisting{ExecPrompt}{
  breakable,
  colback=ExecutionPromptBack,
  colframe=ExecutionPromptFrame,
  colbacktitle=ExecutionTitleBack,
   coltitle=ExecutionPromptTitle,
  boxrule=0.8pt,
  arc=2mm, 
  title=Execution Prompt,
  listing only,
  fontupper=\small\ttfamily,
  left=1mm,right=1mm,top=1mm,bottom=1mm
}

\definecolor{PlanningBack}{RGB}{220,245,255}
\definecolor{PlanningFrame}{RGB}{120,180,200}
\definecolor{PlanningTitleBack}{RGB}{190,230,255}
\definecolor{PlanningTitle}{RGB}{30,50,80}

\newtcblisting{PlanningPrompt}{
  breakable,
  colback=PlanningBack,
  colframe=PlanningFrame,
  colbacktitle=PlanningTitleBack,
  coltitle=PlanningTitle, 
  boxrule=0.8pt,
  arc=2mm, 
  title=Planning Prompt,
  listing only,
  fontupper=\small\ttfamily,
  left=1mm,right=1mm,top=1mm,bottom=1mm
}

\definecolor{EvaluationBack}{RGB}{247,200,195}
\definecolor{EvaluationFrame}{RGB}{247,139,175}
\definecolor{EvaluationTitleBack}{RGB}{247,179,175}
\definecolor{EvaluationTitle}{RGB}{60,30,80}

\newtcblisting{EvaluationPrompt}{
  breakable,
  colback=EvaluationBack,
  colframe=EvaluationFrame,
  colbacktitle=EvaluationTitleBack,
  coltitle=EvaluationTitle, 
  boxrule=0.8pt,
  arc=2mm, 
  title=Evaluation Prompt,
  listing only,
  fontupper=\small\ttfamily,
  left=1mm,right=1mm,top=1mm,bottom=1mm
}

\newtcblisting{RefinePrompt}{
  breakable,
  colback=EvaluationBack,
  colframe=EvaluationFrame,
  colbacktitle=EvaluationTitleBack,
  coltitle=EvaluationTitleBack, 
  boxrule=0.8pt,
  arc=2mm, 
  title=Refined Segmentation Prompt,
  listing only,
  fontupper=\small\ttfamily,
  left=1mm,right=1mm,top=1mm,bottom=1mm
}

\definecolor{SummarizerBack}{RGB}{248,220,250}   
\definecolor{SummarizerFrame}{RGB}{224,160,190}  
\definecolor{SummarizerTitleBack}{RGB}{240,160,240} 
\definecolor{SummarizerTitle}{RGB}{72,34,64}     

\newtcblisting{SummarizePrompt}{
  breakable,
  colback=SummarizerBack,
  colframe=SummarizerFrame,
  colbacktitle=SummarizerTitleBack,
  coltitle=SummarizerTitle, 
  boxrule=0.8pt,
  arc=2mm, 
  title=Summarize \& Personalization Prompt,
  listing only,
  fontupper=\small\ttfamily,
  left=1mm,right=1mm,top=1mm,bottom=1mm
}

\newtcblisting{SearchPrompt}{
  breakable,
  colback=SummarizerBack,
  colframe=SummarizerFrame,
  colbacktitle=SummarizerTitleBack,
  coltitle=SummarizerTitle, 
  boxrule=0.8pt,
  arc=2mm, 
  title=Search Summarize Prompt,
  listing only,
  fontupper=\small\ttfamily,
  left=1mm,right=1mm,top=1mm,bottom=1mm
} 

\theoremstyle{thmstyleone}%
\theoremstyle{thmstyletwo}%

\theoremstyle{thmstylethree}%

\newcommand{\ie}{\textit{i.e., }}
\newcommand{\eg}{\textit{e.g., }}

\begin{document}
\title{GenCellAgent: \textbf{Gen}eralizable, Training-Free \textbf{Cell}ular Image Segmentation via Large Language Model \textbf{Agent}s}

\author*[1]{\fnm{Xi} \sur{Yu}}\email{xyu1@bnl.gov}
\author[2]{\fnm{Yang} \sur{Yang}}\email{yyang@bnl.gov}
\author[3]{\fnm{Qun} \sur{Liu}}\email{qunliu@bnl.gov}
\author[2]{\fnm{Yonghua} \sur{Du}}\email{ydu@bnl.gov}
\author[2]{\fnm{Sean} \sur{McSweeney}}\email{smcsweeney@bnl.gov}
\author*[1]{\fnm{Yuewei} \sur{Lin}}\email{ywlin@bnl.gov}

\affil[1]{\orgdiv{Artificial Intelligence Department}, \orgname{Brookhaven National Laboratory}, \orgaddress{\city{Upton}, \postcode{11973}, \state{NY}, \country{US}}}

\affil[2]{\orgdiv{National Synchrotron Light Source II}, \orgname{Brookhaven National Laboratory}, \orgaddress{\city{Upton}, \postcode{11973}, \state{NY}, \country{US}}}

\affil[3]{\orgdiv{Biology Department}, \orgname{Brookhaven National Laboratory}, \orgaddress{\city{Upton}, \postcode{11973}, \state{NY}, \country{US}}}


\abstract{Cellular image segmentation is essential for quantitative biology yet remains difficult due to heterogeneous modalities, morphological variability, and limited annotations. We present GenCellAgent, a training-free multi-agent framework that orchestrates specialist segmenters and generalist vision-language models via a planner-executor-evaluator loop (choose tool $\rightarrow$ run $\rightarrow$ quality-check) with long-term memory. The system (i) automatically routes images to the best tool, (ii) adapts on the fly using a few reference images when imaging conditions differ from what a tool expects, (iii) supports text-guided segmentation of organelles not covered by existing models, and (iv) commits expert edits to memory, enabling self-evolution and personalized workflows. Across seven cell-segmentation benchmarks spanning diverse microscopy modalities (4,718 images), this routing consistently matches or exceeds the best individual tool on every dataset and outperforms all baselines in overall accuracy. On out-of-distribution organelle data, GenCellAgent substantially outperforms specialist models that were not trained on the target domain, recovering structures that dedicated tools fail to detect. It also segments novel objects such as the Golgi apparatus via iterative text-guided refinement, with light human correction further boosting performance. Together, these capabilities provide a practical path to robust, adaptable cellular image segmentation without retraining, while reducing annotation burden and matching user preferences.}


\maketitle

\section{Introduction}\label{sec1}

Cellular image analysis plays a critical role in modern life sciences by converting complex imaging data into quantitative insight. Within this field, cell and organelle segmentation is particularly crucial, enabling precise measurements, dynamic tracking, and automated, high-throughput analysis. However, performance often degrades when imaging modalities, microscopes, labeling strategies, or sample-preparation protocols differ across laboratories. Retraining models and re-annotating data introduce latency and cost that limit routine use.

The community has responded with powerful but largely specialized solutions for cellular image segmentation.  Classical pipelines based on handcrafted features laid the groundwork~\cite{xie2012biological,wang2015application,humnabadkar2003unsupervised}, but struggle to generalize across modalities. Widely adopted pretrained segmenters such as Cellpose and derivatives~\cite{stringer2021cellpose,pachitariu2022cellpose,stringer2025cellpose3,zhang2025swincell}, SAM~\cite{kirillov2023segment} and its adaptations (like CellSAM~\cite{israel2025cellsam}, $\mu$SAM~\cite{archit2025segment}, MedSAM~\cite{ma2024segment} and CryoSAM~\cite{zhao2024cryosam}), cell simulation as cell segmentation~\cite{jones2024cell}, Nellie~\cite{lefebvre2025nellie}, as well as specialized frameworks like ERNet~\cite{lu2023ernet} and MitoNet~\cite{glancy2023mitonet}, have demonstrated strong performance in cell and nuclei segmentation, while progressively extending their specialization to subcellular structures and diverse microscopy modalities (e.g., medical imaging, cryo-ET). Despite this progress, utility often remains tied to the object classes and domains seen during training, leading to degraded performance on out-of-distribution (OOD) data and novel biological structures. In parallel, advances in large language models have spurred agentic systems that couple reasoning with analysis: Omega~\cite{royer2024omega} enables natural-language execution of image tasks within napari~\cite{chiu2022napari}, BioImage.IO~\cite{lei2024bioimage} provides a conversational interface to models and datasets, and CellAgent~\cite{xiao2024cellagent} demonstrates multi-agent orchestration for non-imaging omics workflows (scRNA-seq). However, current large language model (LLM) agent-based approaches still rely on static planning and typically lack adaptive tool selection, automatic, quality-driven refinement, and long-term memory to learn from lightweight user feedback across sessions.

\begin{figure}[H]
 \centering 
\includegraphics[width=1\textwidth]{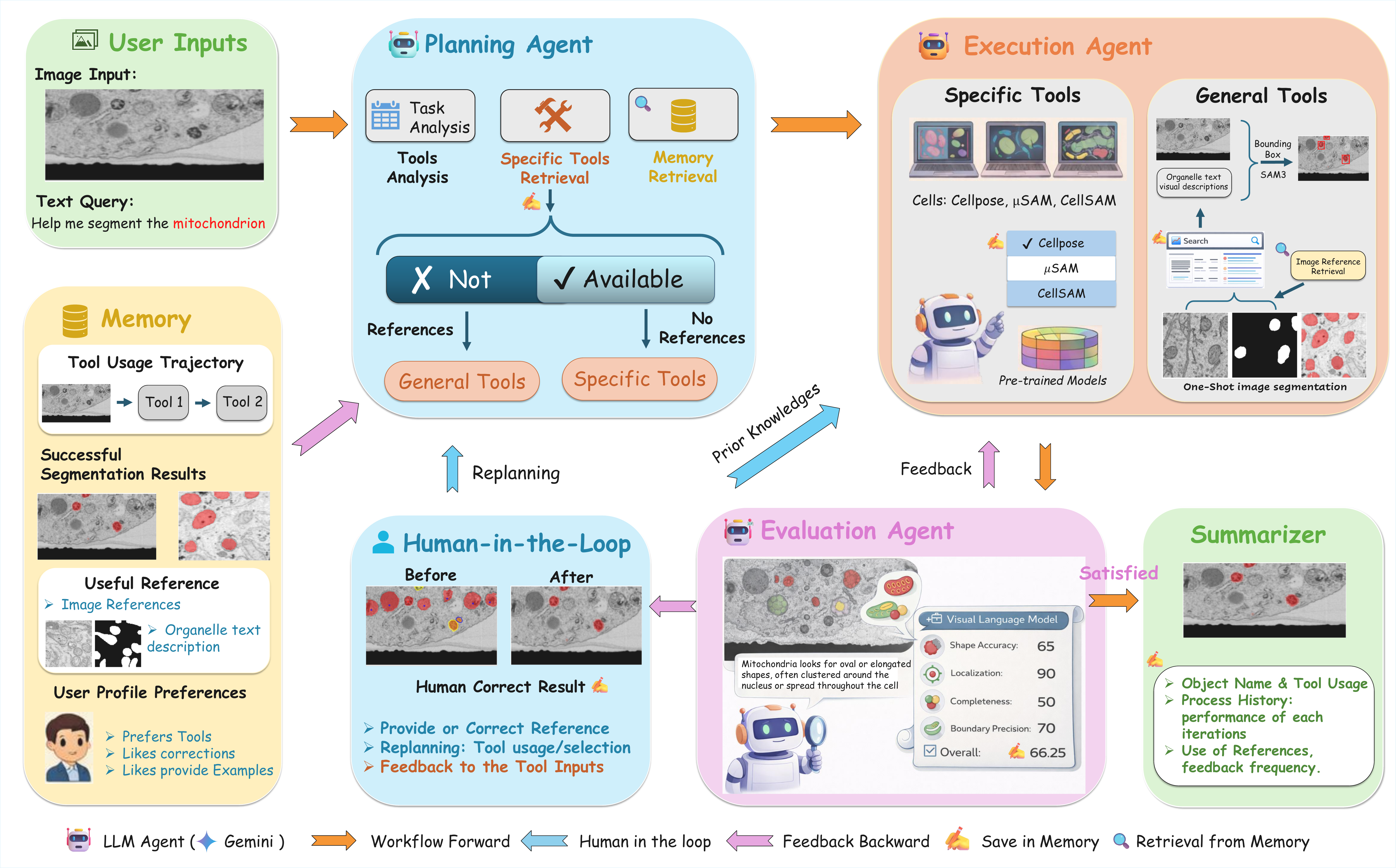}
\caption{Overview of the \textbf{GenCellAgent} framework. The system integrates a Planning Agent, Execution Agent, and Evaluation Agent within a multi-agent
interaction loop, supported by memory and human-in-the-loop (HITL) interaction. \emph{Planning Agent} interprets user queries, retrieves prior knowledge, and
designs workflows. \emph{Execution Agent} runs specialist and generalist tools to produce candidate segmentations. \emph{Evaluation Agent} scores results and provides feedback for refinement. All runs are stored in memory, enabling retrieval-augmented generalization, self-evolving improvement, and personalization in future tasks.}
 \label{fig:overview_figure}
\end{figure}

To address these limitations, we introduce \textbf{GenCellAgent}, a training-free, agentic LLM system for cellular image segmentation, as illustrated in Fig.\ref{fig:overview_figure}. The system comprises three coordinated agents, \ie Planning, Execution, and Evaluation, organized in a structured workflow with iterative feedback and optional human-in-the-loop intervention. During planning, appropriate segmentation tools are selected from both specialized pretrained models and general-purpose generative methods, while execution applies these tools and evaluation provides refinement signals. A memory module further enhances the framework by storing prior segmentation results, tool usage trajectories, and reference examples, which are retrieved to guide future decisions and improve overall performance. It integrates four key capabilities: (i) \textbf{intelligent tool selection}, selecting and configuring state-of-the-art community tools for each image; (ii) \textbf{in-context adaptation}, using a few reference examples to handle shifts in modality or imaging conditions (\eg light microscopy$\rightarrow$electron microscopy) without retraining; (iii) \textbf{text-guided, fully automated segmentation of novel objects} via pretrained vision–language models; (iv) \textbf{human-in-the-loop interaction with memory-driven self-evolution and personalization} wherein experts guide the agent with natural language and intuitive edits, and accepted outputs and preferences are stored in the long-term memory to improve future runs and tailor the balance between automation, time, and quality. Together, these capabilities reduce annotation and retraining burden and make advanced analysis broadly accessible. To our knowledge, this is the first application of LLM agents to cellular image segmentation, linking language-guided reasoning to domain-specific vision tools.

\section{Results}\label{sec:res}
GenCellAgent delivers five capabilities that cover common needs in cellular image segmentation. When dedicated tools exist, it performs style-aware matching to select the best specialist and achieve state-of-the-art accuracy, while under distribution shifts, it escalates to in-context adaptation with retrieved references to refine segmentation  (Capability I). When no tools or references are available, it operates in a fully automated, text-guided mode using morphology descriptions and iterative prompt refinement (Capability II). With human intervention (Capability III), experts can seamlessly edit results, which are incorporated into memory. Over time, this evolving memory accumulates collective knowledge (Capability IV) and adapts to individual users (Capability V), enabling continually improving and personalized workflows.

\subsection{Capability I: Intelligent Tool Selection \& Enhancement -- When Dedicated Tools Exist}\label{sec:s1}
In this section, we demonstrate that GenCellAgent effectively harnesses state-of-the-art segmentation tools by intelligently selecting the most suitable model for diverse datasets, thereby surpassing the performance of any individual tool. We further explore cases where even top-performing segmentation models encounter challenges due to domain shifts.

\subsubsection{Intelligent tool selection}

Our proposed system, GenCellAgent, intelligently selects the most suitable segmentation tool for each image, leveraging complementary strengths from leading models in the research community. As summarized in Table~\ref{specialized_tool}, GenCellAgent orchestrates among six specialized tools spanning cell/nuclei segmentation (Cellpose~\cite{stringer2025cellpose3}, $\mu$SAM~\cite{archit2025segment} and CellSAM~\cite{israel2025cellsam}), mitochondria segmentation (MitoNet~\cite{glancy2023mitonet}), endoplasmic reticulum segmentation (ERNet~\cite{lu2023ernet}) and Golgi apparatus segmentation (an UNet model trained on CellMap dataset by ourselves).

\begin{table}[ht]
\centering
\caption{Specialized Segmentation Tools and Their Target Organelles}
\begin{tabular}{ll}
\hline
\textbf{Tool Name} & \textbf{Target Organelle} \\
\hline
Cellpose & Cell/Nucleus \\
$\mu$SAM & Cell/Nucleus\\
CellSAM &Cell/Nucleus\\
MitoNet & Mitochondria \\
ERNet & ER \\
UNet-Golgi & Golgi \\
\hline
\end{tabular}\label{specialized_tool}
\end{table}

While GenCellAgent orchestrates all six specialized tools, the tool-selection evaluation in this section focuses on the three cell/nucleus segmentation tools, \ie Cellpose~\cite{stringer2025cellpose3}, $\mu$SAM~\cite{archit2025segment} and CellSAM~\cite{israel2025cellsam}, as these cover overlapping target domains and thus require active image-level routing. The remaining tools (MitoNet, ERNet, UNet-Golgi) target distinct organelles and are selected based on the user's task query rather than image-style similarity. Specifically, we quantify tool-selection accuracy by selecting between three representative specialized cell-segmentation tools \ie Cellpose, $\mu$SAM and CellSAM, on seven public cell segmentation benchmarks, LiveCell~\cite{edlund2021livecell} (incucyte HD phase-contrast microscopy images), PlantSeg (Root)~\cite{wolny2020accurate} (confocal laser scanning microscopy images), TissueNet~\cite{greenwald2022whole} (fluorescence and mass spectrometry images), Lizard~\cite{graham2021lizard} (H$\&$E stained histology images), 2018 Data Science Bowl~ \cite{caicedo2019nucleus} (widefield fluorescence), Mouse Brain~\cite{amat2024mesospim} (light-sheet) and Damond ~\cite{damond2024imcdatasets} (imaging mass cytometry). Each specific tool integrated into GenCellAgent is systematically characterized in terms of its supported tasks, the training dataset name, and exemplar sample images where it demonstrates optimal performance. 

\begin{table}[h]
\centering
\caption{Datasets with imaging modalities and corresponding segmentation tool}
\label{tab:modalities}
\small
\begin{tabular}{llcc}
\toprule
\textbf{Dataset} & \textbf{Modality} & \textbf{Best Tool} & \textbf{\# Images} \\
 & & & \textbf{(anchor/test)} \\
\midrule
LIVECell~\cite{edlund2021livecell} & Phase Contrast & Cellpose & 8/1,512 \\
TissueNet~\cite{greenwald2022whole} & Fluorescence \& mass spectrometry & Cellpose & 10/1,311 \\
PlantSeg~\cite{wolny2020accurate} & Confocal & $\mu$SAM & 10/1,408 \\
Lizard~\cite{graham2021lizard} & Brightfield (H\&E) & Cellpose & 5/88 \\
2018 DSB~\cite{caicedo2019nucleus} & Widefield Fluorescence & $\mu$SAM & 11/134 \\
Mouse Brain~\cite{amat2024mesospim} & Light-Sheet & CellSAM & 8/200 \\
Damond~\cite{damond2024imcdatasets} & Imaging Mass Cytometry & CellSAM & 10/65 \\
\midrule
\multicolumn{3}{l}{\textbf{Total}} & \textbf{62/4,718} \\
\bottomrule
\end{tabular}
\end{table}


We quantify tool-selection accuracy as the fraction of test images for which GenCellAgent selects the empirically best segmentation tool. To establish per-image ground truth, we run all available specialist models (Cellpose, $\mu$SAM, and CellSAM) on every test image and identify the tool that achieves the highest segmentation accuracy for that specific image as the oracle best performer. For each test image, GenCellAgent computes its average image-style correlation (details in Appendix~\ref{sec:style}) with anchor images from each candidate dataset, assigns the image to the dataset with the highest similarity, and selects the segmentation tool linked to that dataset. A selection is counted as correct when the chosen tool matches the per-image oracle best performer.

\begin{figure}[hbt!]
 \centering 
 \includegraphics[width=1\textwidth]{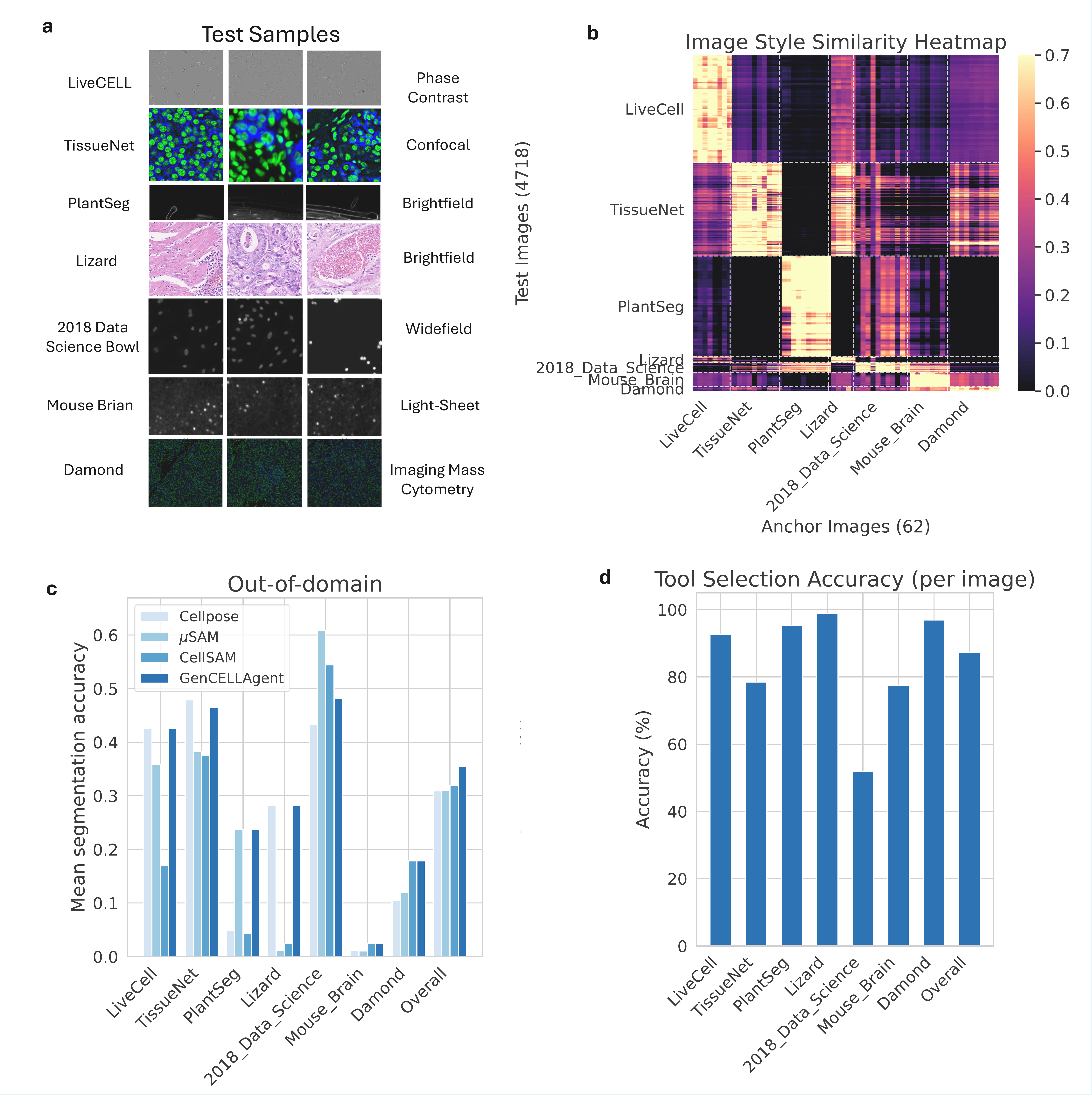}
\caption{\textbf{Intelligent tool selection and performance evaluation}. \textbf{a}, Test image samples from seven public cell segmentation benchmarks spanning diverse microscopy modalities: LIVECell (phase contrast), TissueNet (fluorescence and mass spectrometry), PlantSeg (confocal), Lizard (H$\&$E brightfield), 2018 Data Science Bowl (widefield fluorescence), Mouse Brain (light-sheet), and Damond (imaging mass cytometry). \textbf{b}, Image-style similarity heatmap between 4,718 test images and 62 anchor images, showing modality-based clustering. \textbf{c}, Mean segmentation accuracy comparison across datasets and overall; GenCellAgent achieves comparable performance to the best tool on each individual dataset and delivers the highest overall accuracy. \textbf{d}, Per-image tool-selection accuracy for each dataset and overall, computed against per-image oracle ground truth (the model achieving the highest IoU on each individual image).}
\label{fig:tool_selection}
\end{figure}

In our experimental setup, the evaluation is performed on the full official test sets of all seven benchmarks, totaling 4,718 images (1,512 from LiveCell, 1,408 from PlantSeg, 1,311 from TissueNet, 88 from Lizard, 134 from 2018 Data Science Bowl, 200 from Mouse Brain, and 65 from Damond). A separate anchor set of 62 images (8 from LiveCell, 10 from PlantSeg, 10 from TissueNet, 5 from Lizard, 11 from 2018 Data Science Bowl, 8 from Mouse Brain, and 10 from Damond) is used solely for style-similarity-based model routing and is excluded from all evaluation metrics. As shown in Fig.\ref{fig:tool_selection}b, the style-similarity matrix effectively captures the intrinsic characteristics of different modalities: LiveCell, TissueNet and Lizard cluster together, PlantSeg and 2018 Data Science Bowl form a second cluster, and the other datasets (Mouse Brain and Damond) form additional groupings based on their imaging characteristics, providing a principled basis for tool selection. For example, when a test image originates from TissueNet, the system identifies Lizard as the most visually similar dataset and selects Cellpose as the segmentation tool. Under per-image evaluation, GenCellAgent achieved 87.2\% tool-selection accuracy across the full test set of 4,718 images (Fig.\ref{fig:tool_selection}d), demonstrating that the style-based routing genuinely selects the highest-performing tool for individual images. As illustrated in Fig.\ref{fig:tool_selection}c, GenCellAgent achieves performance comparable to the best-performing tool on each individual dataset; while it consistently outperforms all comparison methods (Cellpose, $\mu$SAM, and CellSAM) in overall segmentation accuracy across the full 4,718-image evaluation (defined by mean segmentation accuracy metric~\cite{everingham2010pascal}), thereby effectively leveraging the complementary strengths of each individual method. 


\subsubsection{Enhancement when dedicated tools underperformed}

Even with correct routing, specialist models can degrade on out-of-distribution (OOD) inputs where imaging conditions differ markedly from their training domains. For example, ERNet~\cite{lu2023ernet} is trained on the dataset of structured illumination microscopy (SIM), a light microscopy, degrades on CellMap’s focused ion beam scanning electron microscopy (FIB-SEM) images~\cite{cellmap2024}; similarly, UNet-Golgi, while effective on its training data, exhibits significant performance drops when applied to FIB-SEM Golgi images with substantial domain shifts. To improve generalization, GenCellAgent integrates SegGPT~\cite{wang2023seggpt}, an in-context segmentation foundation model. When the evaluator’s score falls below a predefined threshold, GenCellAgent retrieves the most similar in-domain image-mask pair as a prompt and applies SegGPT to segment the OOD image. This enables fine-tuning-free adaptation and substantially improves robustness, allowing flexible handling of unseen domains without additional training.

\begin{figure}
 \centering 
 \includegraphics[width=\textwidth]{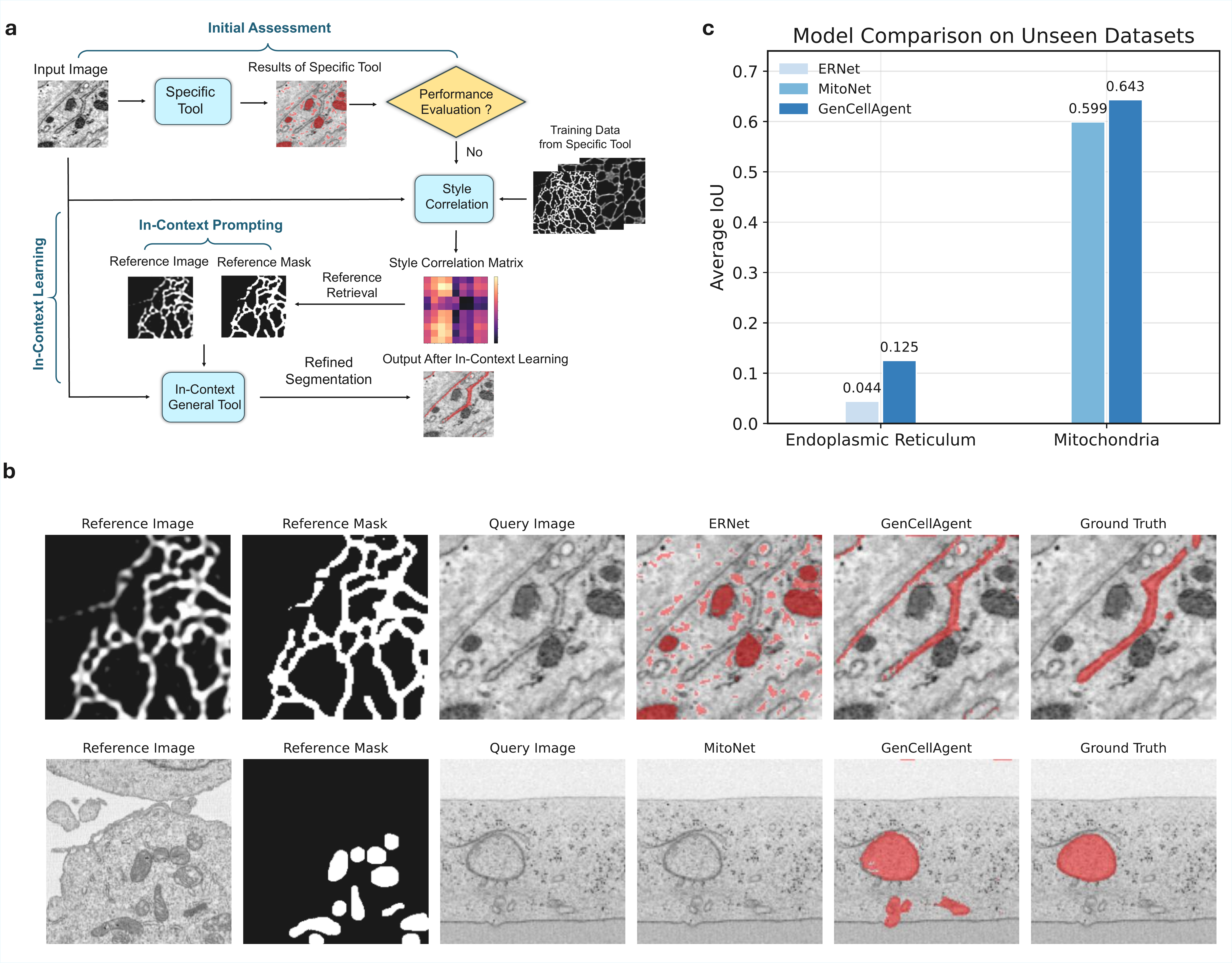}
 \caption{\textbf{Enhancing specialist tools with in-context learning for out-of-distribution segmentation}. 
\textbf{a}, Workflow of the proposed enhancement pipeline. 
\textbf{b}, Qualitative results on unseen datasets.
\textbf{c}, Quantitative comparison.}
 \label{fig:improve_exist_tool}
\end{figure}

Fig.~\ref{fig:improve_exist_tool}a outlines the enhancement workflow. First, the selected specialist tool (e.g., ERNet) generates an initial segmentation. If the evaluation agent determines that the result falls below a predefined quality threshold, GenCellAgent performs reference retrieval, selecting the most style-similar training image for the chosen tool. This image, along with its ground-truth mask, is provided to SegGPT as an in-context prompt, yielding a refined, domain-adapted segmentation. This workflow couples the precision of specialist tools with the adaptability of an in-context segmentation model: when the specialist suffices, the enhancement is skipped; in challenging or OOD cases, SegGPT provides domain-aware improvements.

We evaluated this enhancement on the cellmap datasets~\cite{cellmap2024}, 
both outside the training sets of ERNet~\cite{lu2023ernet} and MitoNet~\cite{glancy2023mitonet}. Note that all evaluations in this section are conducted under a semantic segmentation formulation: instance outputs from tools such as MitoNet are converted to binary foreground masks and evaluated using pixel-level metrics (IoU). This unified protocol ensures fair comparison across all methods, including those that do not produce instance masks. For ER and Golgi, semantic segmentation is the natural formulation due to their continuous morphology; for mitochondria, we focus on foreground region delineation rather than object-level counting. We set the evaluator score threshold to 40; scores below this value trigger the in-context pipeline. Fig.\ref{fig:improve_exist_tool}(b) compares qualitative results: ERNet failed to segment ER structures, producing mainly false positives, whereas GenCellAgent successfully recovered most ER regions with only a few elongated objects misclassified. Similarly, MitoNet did not detect any mitochondrial structures, while GenCellAgent segmented the complete object correctly, with only a few small regions misclassified. The corresponding performance bars in Fig.\ref{fig:improve_exist_tool}c measured by average IoU further highlight GenCellAgent’s substantial improvements over both ERNet and MitoNet, underscoring its robustness in out-of-distribution settings. 

\subsection{Capability II: Fully Automated Segmentation -- Novel Objects Without Dedicated Tools}\label{sec:s2}


This capability addresses the challenging scenario of segmenting novel target objects, for which no object-specific tools, annotated datasets, or prior entries in system memory are available. In this setting, segmentation must rely entirely on textual visual descriptions curated by the execution agent from the literature and web. Such descriptions are often detailed and compositional, involving attributes such as shape, texture, spatial configuration, and contextual relationships. Accurately grounding rich textual information in images remains challenging for models designed for segmentation, such as SAM 3~\cite{carion2025sam}. Although these models exhibit strong mask generation capabilities, this does not necessarily translate into effective text-guided visual reasoning. To address this challenge, we employ an agentic SAM~3~\cite{carion2025sam} pipeline that explicitly separates semantic grounding from mask prediction. Given the target image and a segment prompt describing the object, Gemini~3 first interprets the textual visual description and localizes the most relevant object regions by predicting bounding boxes. These bounding boxes are then used as prompts for SAM~3 to generate the final segmentation masks. This design is motivated by the complementary strengths of the two models: Gemini~3 is highly effective at understanding complex visual-language descriptions and translating them into spatially grounded object hypotheses, whereas SAM~3 excels at precise segmentation once provided with reliable prompts. By combining them in an agentic manner, the system is able to handle complex text-conditioned segmentation tasks more robustly than relying on a segmentation model’s limited textual understanding alone.

Within this framework, the execution agent $\mathcal{A}_{\mathrm{exe}}$ first queries Gemini~3 to produce candidate object locations and then invokes SAM~3 to obtain the corresponding masks. The evaluation agent $\mathcal{A}_{\mathrm{eval}}$ subsequently assesses the segmentation quality and returns feedback. Guided by this feedback, $\mathcal{A}_{\mathrm{exe}}$ refines the segment prompt and re-invokes the Gemini~3--SAM~3 pipeline, thereby enabling iterative improvement. The process terminates after a fixed number of iterations (e.g., 5). At each iteration, the current segmentation result is evaluated, and the best mask is selected based on the highest evaluation score across all iterations. (see Section~\ref{sec:iterative}).

We evaluate this fully automated workflow on three organelle types from the CellMap validation dataset~\cite{cellmap2024}: 278 mitochondria images (128$\times$128, 8$nm$), 200 endoplasmic reticulum images (128$\times$128, 8$nm$), and Golgi images at 128$\times$128 resolution with three voxel sizes, including 200 images at 8$nm$, 110 images at 16$nm$, and 64 images at 32$nm$. All samples are drawn from the validation-split 2D benchmark set (see Appendix~\ref{sec:imggen} for dataset construction details). Importantly, no organelle-specific training data is used; the results are obtained solely from textual task descriptions and iterative feedback.

\begin{figure}[H]
 \centering 
\includegraphics[width=1\textwidth]{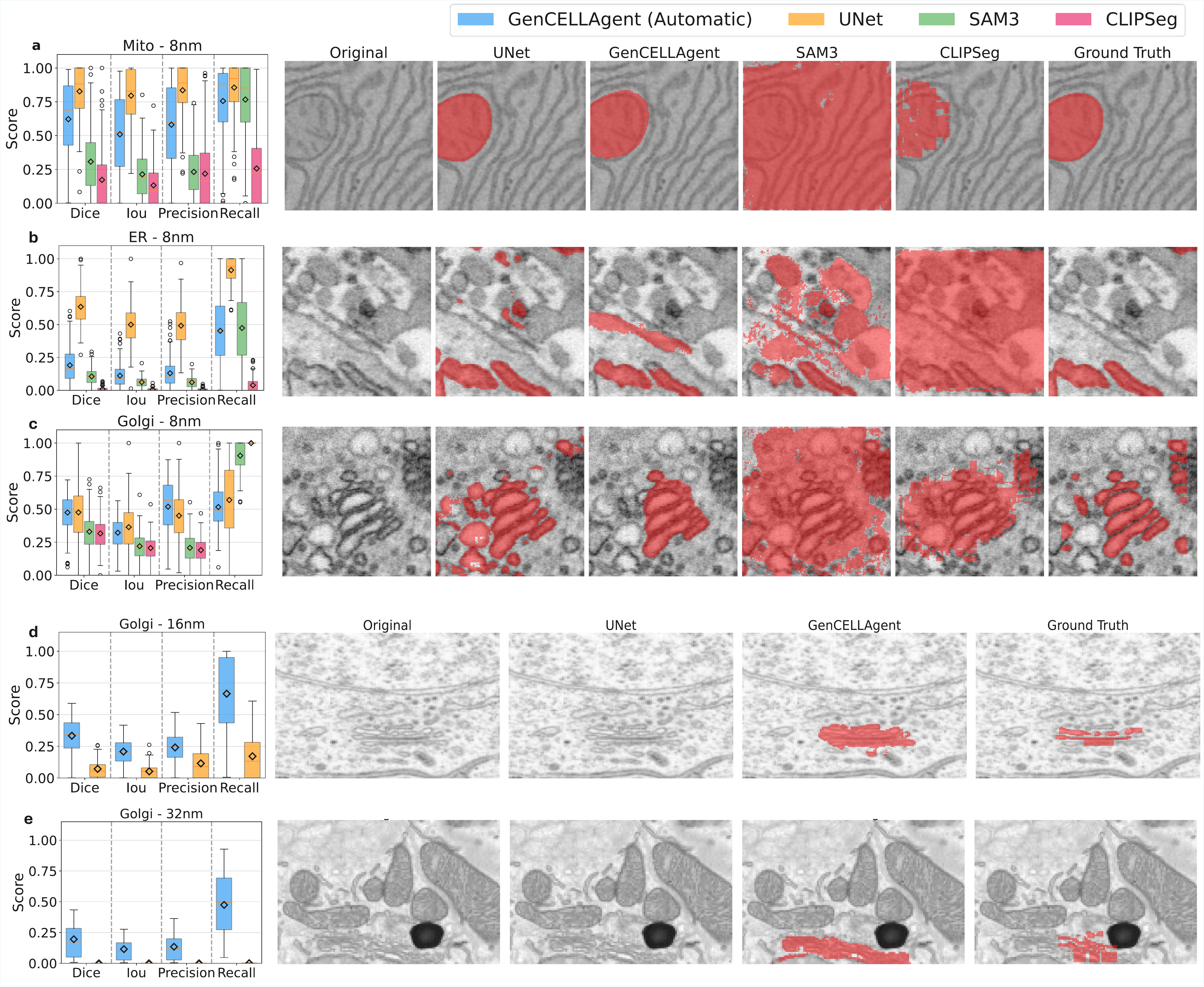}
\caption{Quantitative and qualitative comparison of segmentation methods under in-domain and out-of-domain settings. For each panel, the left shows boxplots of Dice, IoU, Precision, and Recall, and the right shows a representative input image, model predictions, and the ground-truth mask. \textbf{a}, Mito, 8 nm, \textbf{b}, ER, 8 nm, and \textbf{c} Golgi, 8 nm compare all methods: UNet is trained on the corresponding target domain, whereas GenCELLAgent, SAM3, and CLIPSeg are evaluated zero-shot. \textbf{d}, Golgi, 16 nm, and \textbf{e}, Golgi, 32 nm show out-of-domain evaluation, where UNet is trained on Golgi 8 nm and tested at 16 nm and 32 nm, respectively, while GenCELLAgent is applied zero-shot.} 
 \label{fig:fullauto}
\end{figure}

Figure~\ref{fig:fullauto}a presents quantitative results on the mitochondria, ER, and Golgi datasets, all acquired at 8$nm$ resolution. The top row benchmarks two zero-shot baselines, \ie SAM3~\cite{carion2025sam} and CLIPSeg~\cite{luddecke2022image}, together with a supervised in-domain model, UNet, where both training and testing are conducted on 8$nm$ images. In its fully automated segmentation mode, GenCellAgent, powered by the agentic SAM3 model, consistently and substantially outperforms the standalone zero-shot baselines across the main segmentation metrics on all three organelle datasets. Notably, despite requiring no task-specific training, GenCellAgent still delivers performance on the Golgi dataset that is comparable to the supervised model, underscoring the effectiveness of the proposed training-free framework.

To better reflect practical deployment conditions, we further evaluated an out-of-domain setting using the supervised UNet. Specifically, the UNet model is trained on the 8$nm$ Golgi dataset and tested on the  16$nm$ and 32$nm$ Golgi datasets. When evaluated under resolution shifts to 16$nm$ (Figure~\ref{fig:fullauto}d) and 32$nm$ (Figure~\ref{fig:fullauto}e), the U-Net degrades substantially (e.g., Golgi performance almost collapses at 32$nm$), while GenCellAgent (training-free) remains significantly more robust across these resolution shifts, highlighting the practical value of our training-free approach. In contrast, GenCellAgent, operating without any additional training, remains substantially more robust across both 16$nm$ and 32$nm$ settings. This robustness under domain shift is particularly important in real-world biological imaging, where acquisition conditions, imaging resolutions, and sample characteristics often vary across experiments and laboratories. In this context, although organelle-specific supervised models can be incorporated when available (for example, a Golgi-specific U-Net), biological researchers frequently encounter previously uncharacterized structures, atypical disease morphologies, or uncommon reporter patterns for which no pre-trained model exists. Capability II therefore serves as an important zero-shot bridge. By leveraging object description text prompts, GenCellAgent can automatically produce reasonable initial masks, which users can then refine efficiently through human-in-the-loop interaction. This workflow can substantially reduce annotation effort compared with manual segmentation from scratch. 


\subsection{Capability III: Human-in-the-Loop}\label{sec:s4}
We integrated a human-in-the-loop (HITL) mechanism to effectively leverage domain expertise with minimal human effort. This approach allows human experts to guide and refine GenCellAgent's performance at any stage. The interaction is facilitated through a user-friendly graphical user interface (GUI) that enables users to select their preferred tools or correct segmentation errors. This seamless collaboration between the expert and the GenCellAgent ensures that the final results are both accurate and reliable, incorporating the nuanced judgment that only a domain expert can provide.
\begin{figure}[H]
 \centering 
 \includegraphics[width=\textwidth]{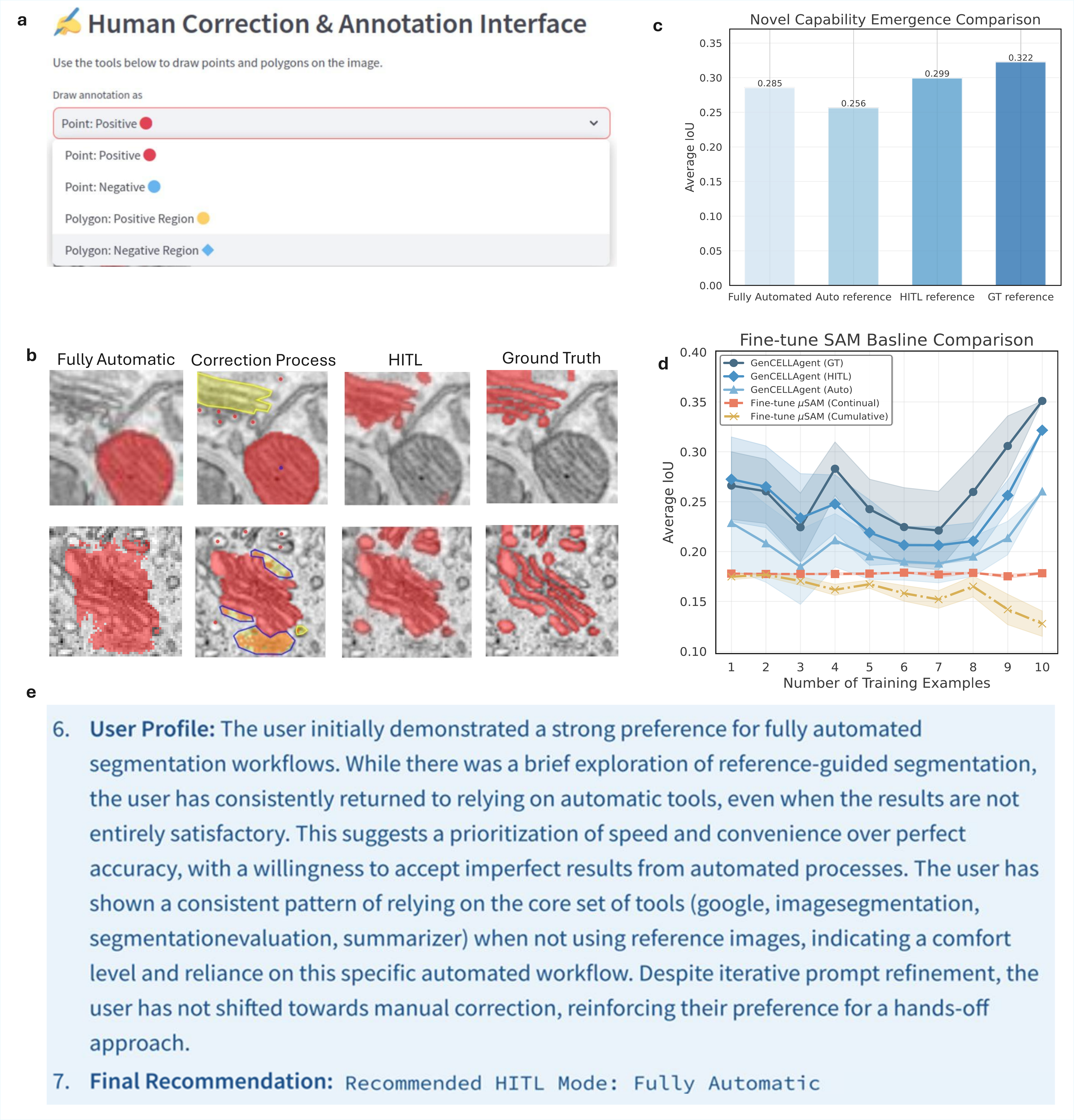}
\caption{\textbf{Human correction, capability emergence, progressive reference learning, and personalized HITL recommendation in GenCellAgent}. \textbf{a}, Human correction interface with four prompt types. \textbf{b}, Segmentation comparison with human correction and ground truth. \textbf{c}, Novel capability emergence across reference scenarios. \textbf{d}, Golgi segmentation performance comparison between GenCELLAgent and fine-tuned $\mu$SAM baselines in a continuous learning setting}. \textbf{e}, Personalized user profiling and recommendation.
 \label{fig:HITL_self_evolution}
\end{figure}

To make human correction simple and effective, we developed a minimal human correction and annotation tool that requires only minimal user input. As illustrated in Fig.\ref{fig:HITL_self_evolution}a, the graphical interface allows users to select different annotation modes. Positive points can quickly trigger detection of an entire object using the SAM-based model, whereas negative points remove an entire object in a single step, allowing large-scale corrections to be performed efficiently. For more complex cases, positive and negative polygons provide the flexibility to add or exclude irregular or difficult-to-detect regions. By combining object-level operations (points) with region-level editing (polygons), this lightweight interface achieves both efficiency through fast, one-click object-level corrections and effectiveness through precise adjustments in challenging or ambiguous regions. This design significantly reduces annotation workload while ensuring high-quality segmentation outcomes. Fig.\ref{fig:HITL_self_evolution}b demonstrates how human corrections improve segmentation quality. Compared to the fully automatic output, the incorporation of human correction leads to refined human-in-the-loop (HITL) results that more closely match the ground truth. This highlights the importance of minimal, targeted human input for substantially improving automated segmentation accuracy. Additionally, humans can intervene in the process and select their preferred tool, more details are described in the Appendix~Fig.\ref{sec:appendix:HITL}.

\subsection{Capability IV: Memory-Driven Self-evolution}\label{sec:s3}

\noindent The self‑evolution of GenCellAgent unfolds across two levels: (1) \textbf{novel capability emergence}, whereby new segmentation tasks previously untenable become feasible; and (2) \textbf{progressive  performance enhancement}, where proficiency improves as experience accumulates.

\subsubsection{Novel Capability Emergence}\label{sec:novel}
Initially, GenCellAgent may lack certain task-specific segmentation capabilities, \eg \textit{``segment golgi''}. In such cases, it first relies on a fully automated way (Section~\ref{sec:s2}) to produce preliminary masks, which are then refined through human correction (Section~\ref{sec:s4}). The resulting image-mask pairs are stored in long-term memory, enabling the system to retrieve them later as reference examples. By leveraging these examples, GenCellAgent can apply in-context learning to perform the \textit{``segment golgi''} task in subsequent cases. In this way, GenCellAgent demonstrates a form of self-evolution, where novel capabilities can emerge even without a dedicated tool for a new task. By combining automatic outputs or human-corrected results with long-term memory, it can leverage in-context learning in a few-shot manner to acquire these new abilities. 

To evaluate this emergent capability, we use a small Golgi dataset of 10 images from cellmap~\cite{cellmap2024} datasets and apply ten-fold cross-validation. In each fold, one image processed either by fully automated segmentation or with human correction is treated as the reference, imitating a \textit{previously processed image}. The remaining nine images are segmented using in-context learning with the reference, testing whether GenCellAgent can learn a new task (here, \textit{golgi segmentation}) from a single example. We evaluated self-evolution under three reference scenarios: fully automated (FA), human correction (HITL), and ground truth (GT).


Fig.~\ref{fig:HITL_self_evolution}c summarizes the average IoU across all scenarios. The fully automated pipeline provides a reasonable baseline, while self-evolution with an automatically generated reference achieves comparable performance, showing that GenCellAgent can transfer its capability to the \textit{segment golgi} task. In practice, once an image has been processed automatically, that result can serve as a reference for new images through in-context learning, avoiding the need to rerun the full and time-consuming automated pipeline. When the reference is further improved by minimal human correction or replaced with ground truth, the self-evolution performance increases further and surpasses the initial automated baseline. This demonstrates that GenCellAgent can learn effectively from higher-quality examples, allowing it not only to acquire new task capabilities but also to improve performance through human expertise.

\subsubsection{Progressive Performance Improvement}\label{sec:progressive}

As GenCellAgent processes more images for the Golgi segmentation task, its performance is expected to improve through increasingly effective in-context referencing. To simulate this continuous learning setting, we progressively increased the number of previously processed images available as reference examples. Specifically, we first generate a random ordering of 10 images to represent the sequence in which labeled samples become available over time. As each image arrives, it is added to the reference pool. Evaluation is then performed on a separate test set of 200 Golgi images that excludes these 10 reference images. Given a set of candidate reference images, we computed a style-similarity score between each test image and every candidate in the reference pool, and selected the most similar image as the reference for in-context prompting

We evaluated the same three scenarios as in Section~\ref{sec:novel}: fully automated, minimally human-corrected, and ground-truth reference. We compared GenCellAgent with SAM-based baselines fine-tuned on 10 training images, and evaluated all methods on the complete CellMap Golgi validation set. The baselines were fine-tuned under two training schemes: (i) \emph{continual fine-tuning}, in which the model is updated sequentially as new examples become available, and (ii) \emph{cumulative fine-tuning}, in which the model is trained on all previously seen examples. These baselines represent standard strategies for adapting SAM-style models in incremental or streaming settings and provide a direct comparison with our style-similarity-based reference selection strategy. To assess robustness, we report results over five different random orderings of the 10 training images.

As shown in Fig.\ref{fig:HITL_self_evolution}d, both SAM-based fine-tuning strategies show limited improvement as additional training examples are introduced. The \textit{continual fine-tuning} baseline remains nearly flat, with mean IoU staying around $0.17$--$0.18$ across $1$--$10$ training examples. The \textit{cumulative fine-tuning} baseline performs even worse, gradually decreasing from approximately $0.17$ to $0.13$ as more examples are added, suggesting instability and limited benefit from naive data accumulation.

In contrast, GenCELLAgent improves as additional examples become available, reaching a mean IoU of approximately $0.35$ with ground-truth annotations, $0.32$ with minimally human-corrected annotations, and $0.26$ with automatically generated labels when 10 examples are available. These results demonstrate that style-similarity-based in-context example selection is substantially more effective than SAM-based fine-tuning strategies for continuous learning in low-data settings.

\subsection{Capability V: Personalized Operation}\label{sec:s5}
Users may have different preferences depending on their expertise and goals. Novice users often prefer fully automated workflows that require minimal input, whereas experienced users may value greater control for fine-grained refinement. As illustrated in Appendix Table.\ref{tab:runtime_memory}, the computational time required varies significantly across different operational modes. To accommodate these diverse needs, GenCellAgent provides flexible personalization, enabling users to balance the level of human-in-the-loop (HITL) interaction against both speed and accuracy. 

When GenCellAgent is executed across multiple sessions, the system records and integrates behavioral information to capture the user’s long-term interaction style. By aggregating data across runs, it can identify stable trends, such as a consistent preference for minimal intervention or reliance on fine-grained manual refinement. Combining short-term behavior with long-term patterns, the system adapts in real time while refining its recommendations over time, ultimately determining a user-preferred HITL level \(\ell_{\text{HITL}}\) from three modes: fully automatic, reference-guided, or human interactive. Fig.\ref{fig:HITL_self_evolution}e shows an example of personalized profiling and recommendation, where the system infers a preference for fully automated workflows and recommends the corresponding HITL mode. See complete user profile recommendation in Appendix~Fig.\ref{GUI_screens_2}. Based on this user preference, GenCellAgent can replan and generate execution strategies tailored to individual needs, dynamically aligning with both task requirements and long-term user behavior to enhance usability and efficiency.

\subsection{Independent validation of the Evaluation Agent}\label{sec:s6}

\begin{figure}[H]
 \centering 
\includegraphics[width=1\textwidth]{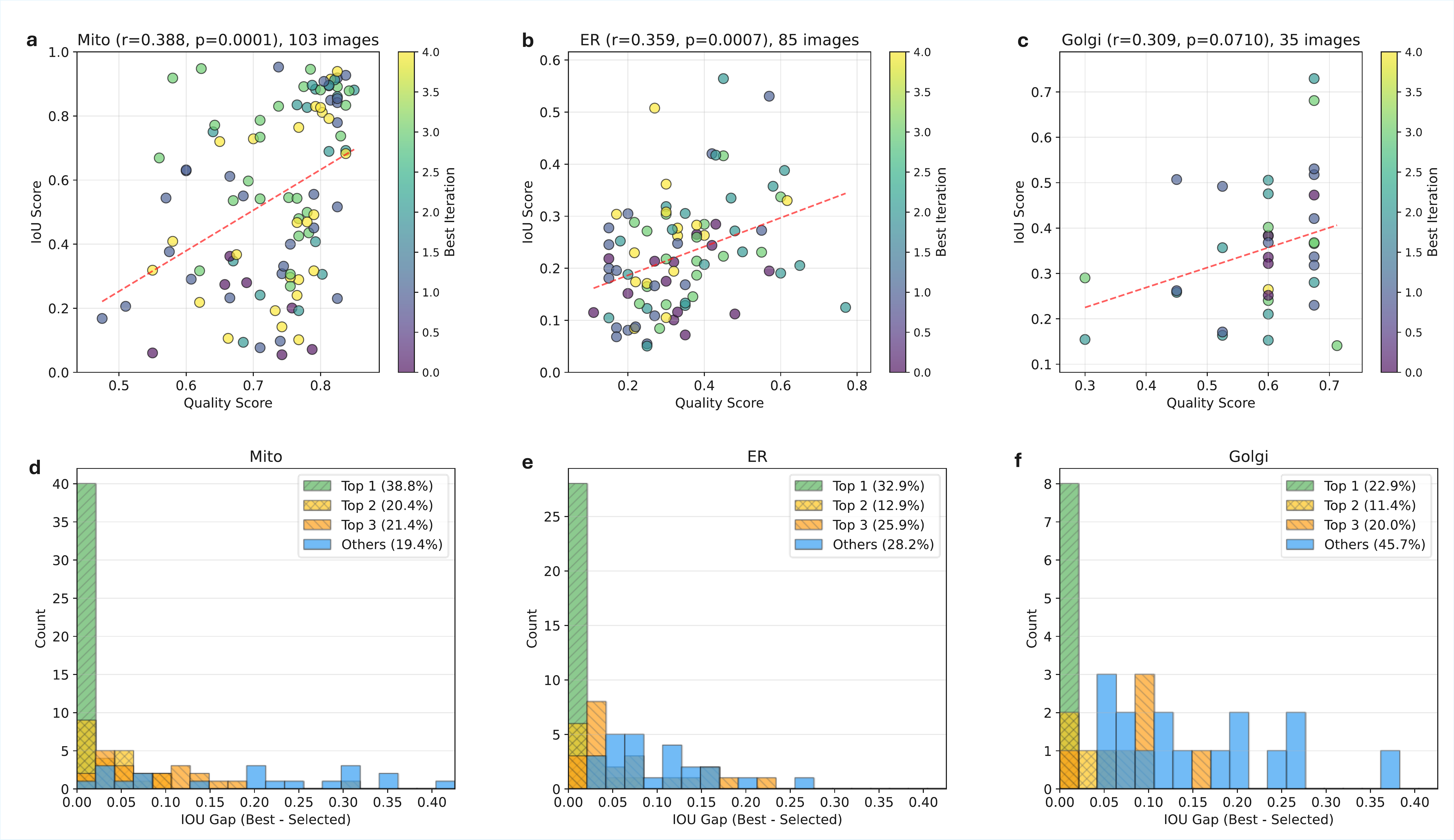}
\caption{Relationship between quality score and segmentation performance across Mito, ER, and Golgi images. \textbf{a-c}: scatter plots show the correlation between quality score and IoU score for each organelle, with points colored by the iteration that produced the best result and red dashed lines indicating linear trends. \textbf{d-f}: histograms of IoU gap between the best-achieved result and the selected result, grouped by whether the best mask was obtained in the top 1, top 2, and top 3.}
 \label{fig:correlation}
\end{figure}

To independently validate the Evaluation Agent, we assessed it from the training-split 2D evaluation set (see Appendix~\ref{sec:imggen} for dataset construction details) in  two perspectives. First, we measured the Pearson correlation between the agent's predicted quality score and the IoU computed against ground-truth annotations, as shown in the first row of Fig.~\ref{fig:correlation}. Across organelles, the Evaluation Agent showed positive correlations with true segmentation performance: mitochondria (r$=$0.388, p$=$0.0001, 103 images), ER (r$=$0.359, p$=$0.0007, 85 images), and Golgi (r$=$0.309, p$=$0.071, 35 images). These results indicate statistically significant moderate correlations for mitochondria and ER, and a weaker positive trend for Golgi that did not reach statistical significance. Overall, the Evaluation Agent captures a quality signal that is positively associated with true segmentation performance, although the strength of this association varies across organelles.

Second, we evaluated selection accuracy in the automatic iterative refinement pipeline. After five iterations, the Evaluation Agent selected the best candidate based solely on its predicted quality score, and we assessed whether this corresponded to the true Top-1, Top-2, or Top-3 segmentation ranked by ground-truth IoU. As shown in the second row of Fig.~\ref{fig:correlation}, the agent selected the true Top-1 candidate in 38.8\% of mitochondria cases, 32.9\% of ER cases, and 22.9\% of Golgi cases. Top-3 accuracy increased to 80.6\% for mitochondria, 71.7\% for ER, and 54.3\% for Golgi. The IoU gap between the best and selected candidates was generally small, particularly for mitochondria, indicating that even when the agent did not identify the single best result, the selected segmentation was often close in quality to the optimum. Overall, these findings suggest that the Evaluation Agent provides a useful, although imperfect, quality signal and ranking mechanism, with stronger performance for structurally distinctive organelles such as mitochondria.

\section{Discussion}\label{sec_discussion}

We present GenCellAgent, a training-free, self-evolving, multi-LLM agent framework for cellular image segmentation that lowers the task barriers by orchestrating specialist segmenters together with generalist vision–language models (VLMs) in a coordinated, multi-agent planning–execution–evaluation loop. Specifically, it (i) selects the most suitable specialist when it is available, (ii) escalates to in-context adaptation when distribution shifts degrade performance, (iii) segments novel targets fully automatically using text-guided prompts with iterative, evaluator-driven refinement when no tool or labels exist, (iv) incorporates human-in-the-loop (HITL) edits via a lightweight correction GUI to enable expert oversight, (v) personalizes workflows over time by remembering accepted outputs and user preferences. This layered strategy transforms heterogeneous community resources into a coherent method that generalizes without retraining while providing intuitive human-in-the-loop interfaces that efficiently integrate expert feedback and guidance, thereby ensuring adaptability to specific research needs.

\textbf{Practical guidance for laboratories.} (1) \textit{Rely on the system to select the tool first.} The planner uses style similarity to select an effective specialist for each image, even on out-of-distribution data, override only if users are confident about a specific tool. (2) \textit{For novel targets, iterate.} Run $\sim3$–$5$ evaluator-guided rounds with light test-time scaling; quality typically improves and stabilizes. (3) \textit{Invest early in minimal edits.} A few point/polygon corrections create high-value references that materially boost subsequent in-context runs and reduce future effort. These practices balance accuracy, computation, and human effort, supporting a progressive workflow: automatic routing first, iterative refinement as needed, and early corrections to steer later improvements. In terms of computational cost, the text-guided workflow costs approximately \$0.012 per image in API fees, with end-to-end processing on the order of 35 seconds, making GenCellAgent practical and affordable for routine use (see Appendix~\ref{sec:cost} for detailed token usage and runtime breakdowns).

\begin{figure}[H]
 \centering 
\includegraphics[width=1\textwidth]{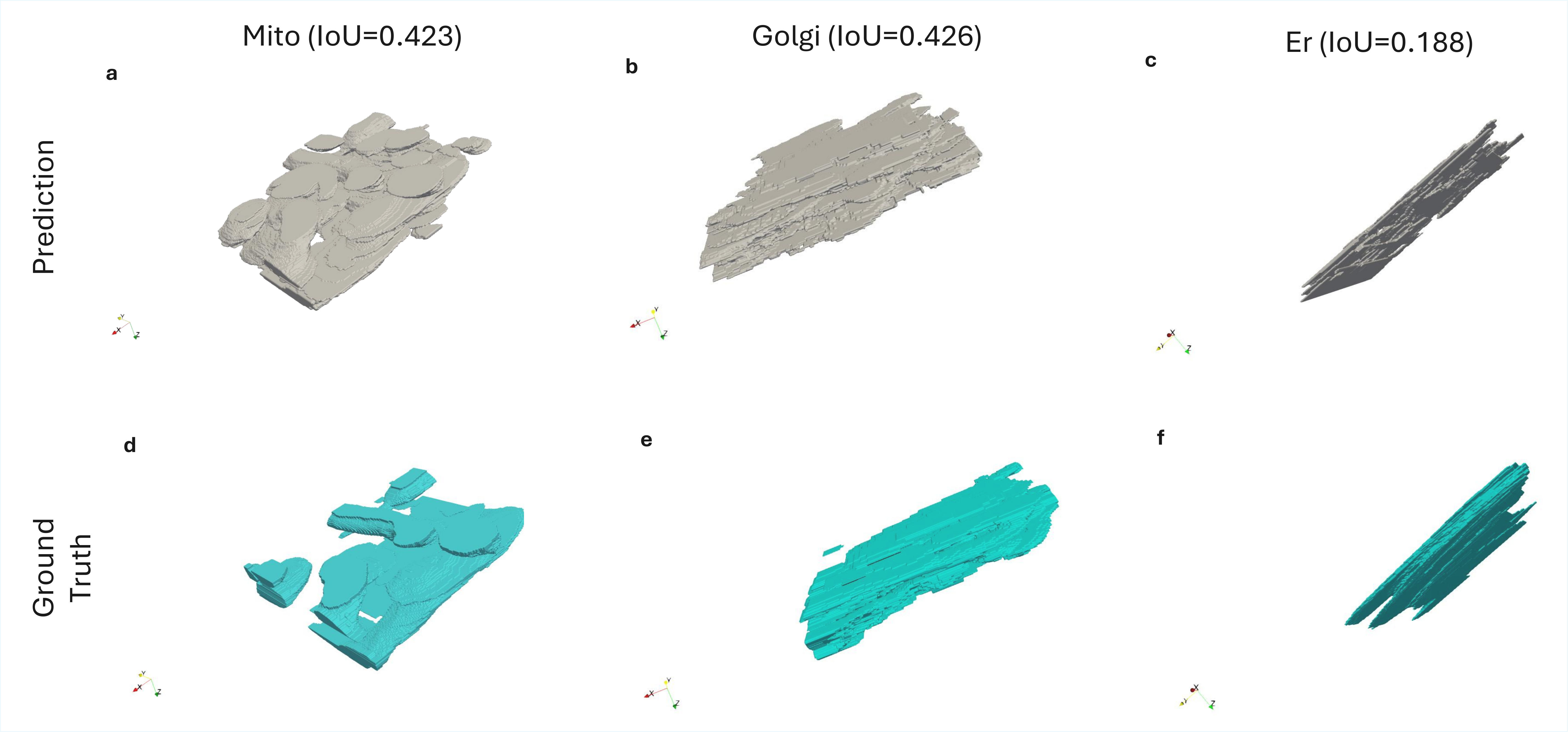}
\caption{Example 3D reconstruction of zero-shot organelle segmentation results using slice-wise 2D inference. \textbf{a-c}: volumetric predictions reconstructed by stacking 2D slice-wise segmentations. \textbf{d-f}: corresponding ground-truth 3D annotations. Each example corresponds to a $512 \times 512 \times 512$ volume at 8$nm$ resolution. For clearer visualization, we select representative samples with relatively high label coverage.}
 \label{fig:3D}
\end{figure}

\textbf{Limitations.} First, the current tools integrated in GenCellAgent only operate on 2D slices. However, volumetric segmentation can be achieved via slice-wise 2D inference and stacking. We experimentally validated this strategy by applying our automatic 2D pipeline to full 3D volumes, achieving IoU scores of 0.423 for mitochondria, 0.426 for Golgi, and 0.188 for ER (see Fig.~\ref{fig:3D}). GenCellAgent itself is model-agnostic and not fundamentally limited to 2D: volumetric models (e.g., 3D U-Net) can be integrated without architectural modification. Second, in fully automated mode, failures often arise from domain gaps in the vision–language segmenter (SAM 3). Because it is trained mostly on natural image–text pairs, it has limited exposure to biological data such as fluorescence contrast, EM texture, or topology-rich descriptions of subcellular structures. As a result, it may miss fine processes, oversmooth thin membranes, or struggle in dense and low-SNR fields. Three, while the Evaluation Agent can serve as a relative scorer, the LLM score is often sensitive to prompt phrasing, and prone to overconfidence or misleading correlations in retrieved descriptions. Finally, the system relies on the quality of retrieved descriptions and references. Poor text or unrepresentative examples can bias prompts and memory, leading to over-fitting to recent but uninformative cases.

\textbf{Future directions.} We plan to strengthen the fully automated path with bioimage-specialized vision–language models, adapted on large biological corpora with morphology-aware objectives, paired with evaluators tuned to topology, boundary continuity, and instance separation. Lightweight adapters or efficient fine-tuning may preserve generality while improving biological fidelity. We also aim to extend beyond single-object targets to multi-object segmentation, higher-dimensional data (3D/4D), instance tracking, and joint segmentation–tracking. While the current slice-wise 2D inference approach provides a practical baseline for volumetric analysis, native 3D segmentation pipelines guided by textual descriptions or 3D voxel-level references could further improve volumetric consistency. These extensions will require temporal consistency metrics and memory mechanisms that store short spatiotemporal exemplars rather than isolated images. Finally, we will deepen ecosystem integration by coupling GenCellAgent with tool registries, model hubs, and the Model Context Protocol (MCP) to streamline registration of new tools, routing across heterogeneous backends, and composition of multi-step analyses. Taken together, these directions position GenCellAgent as a practical bridge between agentic language-model control and domain-specific vision tools, with a path toward broader, training-free bioimage workflows in everyday laboratory settings.

\section{Methods}\label{sec11}
\subsection{Large Language Models (LLM) Agent}\label{sec:LLMagent}
Large Language Model (LLM) agents~\cite{luo2025large} extend the capabilities of standalone LLMs with reasoning, planning, memory~\cite{zhang2024survey}, and tool use~\cite{huang2402understanding}, enabling more autonomous and personalized interactions. Beyond single-agent systems, multi-agent frameworks further enhance problem-solving by coordinating specialized agents through structured communication~\cite{guo2024large}. Recent studies show that such agentic systems can accelerate scientific discovery by automating literature review, hypothesis generation, experimental design, and data analysis~\cite{ifargan2025autonomous, zheng2025automation, xiao2025csr, gridach2025agentic, song2025multiagent}. Several frameworks exemplify this trend, including the ``Agent Laboratory'' for autonomous research execution~\cite{schmidgall2025agent}, SciAgents for hypothesis generation in materials science~\cite{ghafarollahi2025sciagents}, LLM-Duo for dual agent text analysis~\cite{hu2024automating}, and Agent-Based Auto Research for end-to-end scientific workflows~\cite{liu2025vision}. Most recently, CRISPR-GPT~\cite{qu2025crispr} was proposed for automatic gene-editing experiments, and Biomni~\cite{huang2025biomni}, a general-purpose biomedical AI agent designed to autonomously execute a wide spectrum of research tasks across diverse biomedical subfields.

While prior approaches highlight the transformative potential of LLM agents in scientific research, they have largely focused on either general-purpose automation or domain-specific text-based tasks. In contrast, our framework is tailored to \emph{cellular image segmentation}, a domain characterized by unique challenges such as ambiguous boundaries, irregular morphologies, and context-dependent structures. Our system brings together planning, execution, evaluation, memory, and human-in-the-loop interaction within a unified agentic architecture. Distinct from previous LLM agent work, our design combines vision–language modeling with iterative refinement and experiential memory, enabling a self-improving segmentation process that balances automation with biological plausibility

\subsection{Proposed Multiple LLM Agents System}

In the multi-agent system, we deploy three LLM-based agents, the \textbf{Planning Agent} $\mathcal{A}_\mathrm{plan}$, the \textbf{Execution Agent} $\mathcal{A}_\mathrm{exc}$ and the \textbf{Evaluation Agent} $\mathcal{A}_\mathrm{eval}$. Planning and execution agents are implemented as Gemini 2.5 Flash and the evaluation agent is implemented Gemini 3.0 Flash instances. Each agent operates as an independent LLM call with role-specific system prompts.

\subsubsection{Planning Agent}\label{sec:plan}
The Planning Agent $\mathcal{A}_{\mathrm{plan}}$ serves as the \textit{central coordinator} of the system. It first receives the target image $\mathbf{Img}_{\mathrm{tar}}$, the user query $\mathcal{Q}$, and the overall task specification $\mathcal{T}$. Using this input, together with tool descriptions $\mathcal{D}$, the agent formulates an initial segmentation plan. See initial planning prompt in Appendix~\ref{sec:appB:planprompt}. This plan guides the Execution Agent $\mathcal{A}_{\mathrm{exc}}$ in invoking the appropriate tools, which may include search engines, specialized segmentation models, texture-aware segmentation methods, and reference image/mask in-context learning tools.

As the workflow progresses, the Planning Agent does not follow a static strategy. Instead, it continuously revises its plan in response to tool outputs and feedback from the Evaluation Agent $\mathcal{A}_{\mathrm{eval}}$. This dynamic adjustment enables the system to refine its segmentation process and select the most suitable tools at each step. Formally, the planning process can be expressed as:
\begin{equation}
\mathcal{P}_{\mathrm{init}} = \mathcal{A}_{\mathrm{plan}}\!\left(\mathbf{Img}_{\mathrm{tar}}, \mathcal{Q}, \mathcal{T}, \mathcal{D}\right),
\end{equation}
\begin{equation}
\mathcal{P}_{t+1} = \mathcal{A}_{\mathrm{plan}}\!\left(\mathcal{P}_{t},\, s_{t+1}\right),
\end{equation}
where $\mathcal{P}_{\mathrm{init}}$ represents the initial segmentation plan, generated by integrating user intent (captured in the query $\mathcal{Q}$ and task definition $\mathcal{T}$) with available tool capabilities ($\mathcal{D}$). At each subsequent step $t+1$, the Planning Agent updates the plan based on the previous strategy $\mathcal{P}_{t}$ and the newly observed state or feedback signal $s_{t+1}$, which may include tool outputs, intermediate segmentation results, or evaluation feedback.

\subsubsection{Execution Agent}\label{sec:exc} 
The Execution Agent $\mathcal{A}_{\mathrm{exe}}$ orchestrates the use of all tools
within the existing tool pool (see Appendix~\ref{sec:appA} for tool descriptions). These tools
span multiple categories. For information collection and summarization,
$\mathcal{A}_{\mathrm{exe}}$ employs web search, literature search, and memory retrieval
to gather task-specific information $\mathcal{I}_{\mathcal{T}}$. For segmentation, the agent invokes a range of specialized and general-purpose models. Specialized tools include CellPose~\cite{stringer2021cellpose, stringer2025cellpose3}, $\mu$SAM~\cite{archit2025segment}, and CellSAM~\cite{israel2025cellsam} for cell/nuclei segmentation, MitoNet~\cite{glancy2023mitonet} for mitochondria segmentation, ERNet~\cite{lu2023ernet} for endoplasmic reticulum segmentation, and UNet-Golgi for Golgi apparatus segmentation, (see Table~\ref{specialized_tool} for a summary). When a user query specifies the target object, the Planning Agent first narrows the candidate tools to those relevant for that object; when multiple tools cover the same target, style-similarity-based routing selects the best performer for the specific image. General-purpose capabilities are provided by agentic SAM3, a vision-language segmentation model that accepts a target image together with a textual description of the object of interest and outputs the corresponding segmentation mask. In
addition, SegGPT~\cite{wang2023seggpt} serves as an in-context learning tool that
takes a reference image and its mask, infers object features, and segments
corresponding objects in the target image. Finally, interactive refinement is
supported by the annotation and correction tool $\mu$SAM~\cite{archit2025segment}, which enables users to correct or refine segmentation results. The execution prompts are shown in Appendix~\ref{sec:appB:exeprompt}.


\noindent \subsubsection{Evaluation Agent}\label{sec:eval}
The Evaluation Agent $\mathcal{A}_\mathrm{eval}$ is implemented as a Large Vision-Language Model (LVLM), capable of integrating both visual and textual information to assess segmentation results. Its core functions are outlined as follows:

\noindent \textbf{Generation of Evaluation Criteria.}  
Given task-specific information $\mathcal{I}_{\mathcal{T}}$, the evaluation agent $\mathcal{A}_{\mathrm{eval}}$ produces a set of task-aware evaluation criteria:
\begin{equation}
\mathcal{C} = \mathcal{A}_{\mathrm{eval}}(\mathcal{I}_{\mathcal{T}},\, \mathbf{p}_{\mathrm{crit}}),
\end{equation}
where $\mathbf{p}_{\mathrm{crit}}$ is a guiding prompt for criteria generation.

\noindent \textbf{Generation of Evaluation Prompt.}  
Conditioned on the criteria $\mathcal{C}$, the agent then constructs a natural-language evaluation prompt:
\begin{equation}
\mathbf{p}_{\mathrm{eval}} = \mathcal{A}_{\mathrm{eval}}(\mathcal{C},\, \mathbf{p}_{\mathrm{gen}}),
\end{equation}
where $\mathbf{p}_{\mathrm{gen}}$ specifies how to translate the criteria into a structured evaluation instruction.  
Both $\mathcal{C}$ and $\mathbf{p}_{\mathrm{eval}}$ are fixed once generated and reused throughout all evaluations for the specific task ($i.e.$, golgi segmentation). See details in Appendix~\ref{sec:appB:evaprompt} for criteria $\mathcal{C}$ and evaluation prompt $\mathbf{p}_{\mathrm{eval}}$. 

\noindent \textbf{Evaluation of Segmentation Results.}  
For a given segmentation result $\mathbf{Mask}_{\mathrm{seg}}$, the evaluation agent measures its quality according to $\mathcal{C}$ under evaluation prompt $\mathbf{p}_{\mathrm{eval}}$:
\begin{equation}
e_{\mathrm{score}},\; e_{\mathrm{sum}}
= \mathcal{A}_{\mathrm{eval}}(\mathbf{Mask}_{\mathrm{seg}},\, \mathcal{C},\, \mathbf{p}_{\mathrm{eval}}),
\end{equation}
where $e_{\mathrm{score}} \in [0,100]$ denotes a numerical quality score, and $e_{\mathrm{sum}}$ provides a textual justification together with suggestions for refinement.

Finally, both $e_{\mathrm{score}}$ and $e_{\mathrm{sum}}$ are forwarded to the Planning Agent $\mathcal{A}_{\mathrm{plan}}$ to support decision-making. If $\mathcal{A}_{\mathrm{plan}}$ initiates another round of segmentation, the summary $e_{\mathrm{sum}}$ is additionally passed to the segmentation prompt generator to refine prompts for Gemini 3 VLLM to refine the bonding box outputs (see Section~\ref{sec:iterative}).


\subsubsection{Memory}
The memory module $\mathcal{M}$ is designed to systematically store and organize
knowledge accumulated from previous runs. It retains key elements such as
retrieved demonstrations, selected tools, segmentation outputs, evaluation
metrics, and user feedback. By continuously incorporating this historical
context, the memory enriches the system’s knowledge base and improves its
decision-making capabilities, thereby enabling more effective and informed
future operations.

After each run, the summarization tool $\mathcal{S}_{\mathrm{sum}}$ generates a
comprehensive summary $\mathbf{S}$ that encapsulates the execution process
$\mathbf{\Pi}$, the task-specific description $\mathcal{I}_{\mathcal{T}}$, the
target image $\mathbf{Img}_{\mathrm{tar}}$, and any reference image and mask
$\{\mathbf{Img}_{\mathrm{ref}}, \mathbf{Mask}_{\mathrm{ref}}\}$:
\begin{equation}
\mathbf{S} = \mathcal{S}_{\mathrm{sum}}\!\big(
\mathbf{\Pi}, \mathbf{Img}_{\mathrm{tar}}, \mathcal{I}_{\mathcal{T}}, \{\mathbf{Img}_{\mathrm{ref}}, \mathbf{Mask}_{\mathrm{ref}}\}\big),
\end{equation}
where the process trace is represented as
\[
\mathbf{\Pi} = \{(in_m,\, out_m,\, act_m)\}_{m=1}^{M},
\]
with $in_m$, $out_m$, and $act_m$ denoting the input, output, and action of the
$m$-th step, respectively, and $M$ indicating the total number of steps in the run.

The memory after $T$ runs is therefore defined as
\begin{equation}
\mathcal{M} = \{\mathbf{S}_{t}\}_{t=1}^{T},
\end{equation}
where $\mathbf{S}_{t}$ is the summary of the $t$-th run. As $T$ grows,
$\mathcal{M}$ progressively expands to capture richer historical context,
thereby providing increasingly valuable guidance for subsequent tasks. See summarize prompts in Appendix~\ref{sec:appB:sum_personprompt}.


\subsection{Text-guided General Segmentation with Iterative Refinement}\label{sec:iterative}

\begin{figure}[H]
 \centering 
 \includegraphics[width=\textwidth]{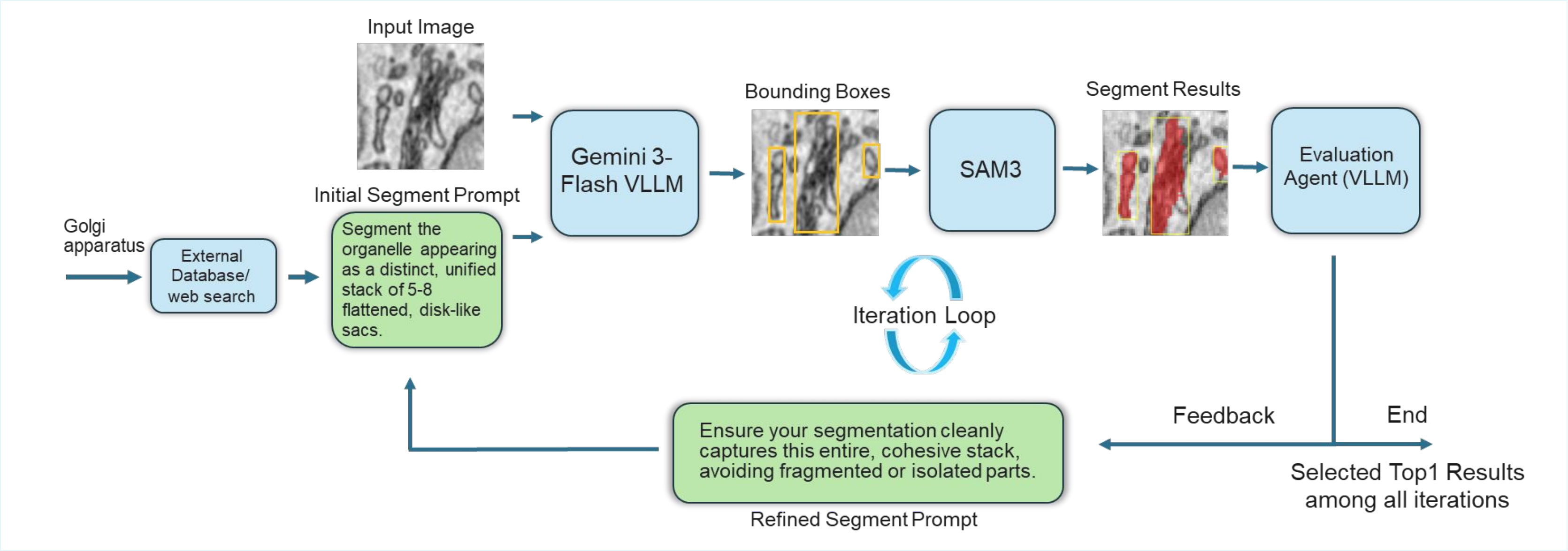}
 \caption{\textbf{Automatic Workflow framework in GenCellAgent}, zoomed in on an area of interest.}
 \label{fig:text_guided}
\end{figure}




Given an input image $\mathbf{Img}_{\mathrm{tar}}$ and a text query \(q\), the goal is to generate a segmentation mask $\mathbf{Mask}_{\mathrm{tar}}$ that matches the target described by the user. As shown in Fig.~\ref{fig:text_guided}, the query is first used to retrieve external web knowledge and construct an initial segmentation prompt \(p_0\). This prompt, together with the image, is then processed by a general segmentation model \(\mathcal{S}_{\mathrm{seg}}=\{G,S\}\), where \(G\) denotes a grounding module that predicts intermediate localization cues and \(S\) denotes a segmentation module that produces the final mask. An evaluation agent further assesses the segmentation result and provides both a quality score and feedback for prompt refinement.

Formally, the process can be written as
\[
p_0 = \mathcal{R}(q), \quad
\mathcal{B}_t = G(\mathbf{Img}_{\mathrm{tar}}, p_t), \quad
\mathbf{Mask}_{\mathrm{tar}}^{\,t} = S(\mathbf{Img}_{\mathrm{tar}}, \mathcal{B}_t),
\]
where \(\mathcal{B}_t\) is the set of predicted bounding boxes at iteration \(t\).

The evaluation agent is defined as
\begin{equation}
e_{\mathrm{score}}^{\,t},\; e_{\mathrm{sum}}^{\,t}
= \mathcal{A}_{\mathrm{eval}}(\mathbf{Mask}_{\mathrm{tar}}^{\,t},\, \mathcal{C},\, \mathbf{p}_{\mathrm{eval}}),
\end{equation}
where \(e_{\mathrm{score}}^{\,t}\) is the segmentation quality score and \(e_{\mathrm{sum}}^{\,t}\) is the textual feedback used to refine the prompt for the next iteration.

To improve segmentation performance, we incorporate a refinement function \(\varphi_{\mathrm{ref}}\) that updates the segmentation prompt. See Appendix~\ref{sec:appB:refsegprompt} for details of the segmentation refinement prompt based on the evaluation feedback. Given the current prompt \(\mathbf{p}_{\mathrm{seg}}^{\,i}\) and evaluation feedback \(\mathbf{e}_{\mathrm{sum}}^{\,i}\), the refinement function generates a single updated prompt for the next iteration:
\begin{equation}
\mathbf{p}_{\mathrm{seg}}^{\,i+1}
= \varphi_{\mathrm{ref}}\!\left(\mathbf{p}_{\mathrm{seg}}^{\,i}, \mathbf{e}_{\mathrm{sum}}^{\,i}\right).
\end{equation}
The updated prompt is then used to produce a new segmentation mask, which is evaluated in the same manner. Across all iterations, the final output mask is selected as the one achieving the highest evaluation score:
\begin{equation}
i^{\star} = \arg\max_i \; e_{\mathrm{score}}^{\,i}, \qquad
\mathbf{Mask}_{\mathrm{tar}}^{\star} = \mathbf{Mask}_{\mathrm{tar}}^{\,i^{\star}}.
\end{equation}

\subsection{In Context Learning}\label{sec:in-context}

In-context learning (ICL) for image segmentation enables pretrained
vision-language models to adapt to new tasks at inference time using only a few
annotated examples, without task-specific retraining. By providing image–mask
pairs (a \emph{reference set}), the model is guided to segment novel objects in
a manner similar to human demonstration. This reduces annotation costs and
improves adaptability, allowing generalization to unseen domains with minimal
supervision.

We adopt SegGPT~\cite{wang2023seggpt} as our ICL segmentation model,
$\mathcal{S}_{\mathrm{ICL}}$. SegGPT is a large-scale visual foundation model
that unifies semantic, instance, and panoptic segmentation by reframing them as
a generalized inpainting problem, enabling transfer to novel object types
without fine-tuning. During inference, the model receives a multi-channel input
consisting of a reference image and its mask, followed by the target image whose
mask is to be predicted. These inputs are tokenized into a sequence of patches
that the Transformer processes to infer the segmentation task from context:
\begin{equation}
\mathbf{Mask}_{\mathrm{tar}} =
\mathcal{S}_{\mathrm{ICL}}\big(\mathbf{Img}_{\mathrm{tar}},
\{\mathbf{Img}_{\mathrm{ref}}, \mathbf{Mask}_{\mathrm{ref}}\}\big).
\end{equation}
Here, the target image $\mathbf{Img}_{\mathrm{tar}}$ and reference pair
$\{\mathbf{Img}_{\mathrm{ref}}, \mathbf{Mask}_{\mathrm{ref}}\}$ are used to
produce the predicted segmentation mask $\mathbf{Mask}_{\mathrm{tar}}$.

In-context learning (ICL) is applied in two main scenarios. First, it can augment existing tools by enhancing their performance when initial results are unsatisfactory. Second, it enables self-evolution, allowing the system to extend its capabilities and tackle segmentation tasks that would otherwise be infeasible.

\subsection{Human In The Loop Interaction}\label{sec:HITL}

Biological image segmentation, particularly at the subcellular level, faces
challenges such as ambiguous boundaries, irregular morphologies, and
context-dependent structural variations. These factors often hinder fully
automated approaches and highlight the importance of human expertise. To address
this, we incorporate a human-in-the-loop (HITL) interaction mechanism that
allows domain experts to intervene strategically at critical decision points,
providing targeted feedback to enhance accuracy.

Within the workflow, users can pause, redirect, or refine steps through short
prompts, offering guidance when automated methods are uncertain. Prior to
finalization, experts may directly edit the segmentation by adding or removing
objects and adjusting boundaries to ensure biological plausibility and
experimental validity. By combining advanced automated models with expert
intervention, our approach produces results that are both accurate and
biologically meaningful. This strategy accelerates analysis while maintaining a
balance between automation and minimal human involvement.


\subsection{Self-Evolving System}

A key limitation in biological image analysis is the static deployment of
workflows—once a segmentation pipeline is built, it rarely benefits from prior
experience. As a result, analyzing similar images often involves redundancy and
inefficiency. To address this, our framework introduces \emph{on-the-fly
experiential learning}, dynamically reusing and evolving from previous runs.

When a user requests a new segmentation task, \eg e.g., \textit{``segment golgi,''}
the Planning Agent $\mathcal{A}_\mathrm{plan}$ queries the memory $\mathcal{M}$
for relevant past workflows. It retrieves procedural blueprints, appearance
descriptors, and previously segmented images of golgi, enabling the
process to be jump-started with proven strategies rather than built from
scratch. These stored examples serve as visual demonstrations, allowing the
system to perform in-context learning even in the absence of specialized
tools, SegGPT can be guided by task-relevant memory entries
(see Section~\ref{sec:in-context}).

To enable retrieval, we construct a query embedding that encodes both the target
image and task description:
\begin{equation}
\mathbf{q} = \{\phi(\mathbf{Img}_{\mathrm{tar}}),\, \tau\},
\end{equation}
where $\mathbf{Img}_{\mathrm{tar}}$ is the target image to be segmented, $\tau$
is the textual task description \eg\textit{segment golgi}, and
$\phi(\cdot)$ is a pre-trained encoder mapping an image into a shared embedding
space.

The Planning Agent $\mathcal{A}_{\mathrm{plan}}$ then queries memory
$\mathcal{M}$ with $\mathbf{q}$. The retrieval operator $\mathcal{R}$ selects
the most relevant reference image/mask example entry:
\begin{equation}
(\mathbf{Img}^*_{\mathrm{ref}},\, \mathbf{Mask}^*_{\mathrm{ref}})
= \mathcal{R}(\mathcal{M}, \mathbf{q})
= \underset{(\mathbf{Img}, \mathbf{Mask}) \in \mathcal{M}}{\arg\max}
\ \mathrm{sim}\!\left(\phi(\mathbf{Img}, \mathbf{Mask}),\, \mathbf{q}\right),
\end{equation}
where $(\mathbf{Img}^*_{\mathrm{ref}}, \mathbf{Mask}^*_{\mathrm{ref}})$ is the
retrieved reference pair, and $\mathrm{sim}(\cdot,\cdot)$ denotes the image style similarity correlation function (see Appendix~\ref{sec:style} Eq.\ref{eq:overall_style_similarity}) used for retrieval similar image in the embedding space.

The in-context learning segmentation model $\mathcal{S}_{\mathrm{ICL}}$ then
uses the retrieved example as a demonstration to predict the segmentation mask
for the target image:
\begin{equation}
\mathbf{Mask}_{\mathrm{tar}}
= \mathcal{S}_{\mathrm{ICL}}\!\left(\mathbf{Img}_{\mathrm{tar}},\,
(\mathbf{Img}^*_{\mathrm{ref}},\, \mathbf{Mask}^*_{\mathrm{ref}})\right),
\end{equation}
where $\mathbf{Mask}_{\mathrm{tar}}$ is the predicted mask for the target image,
conditioned on both the target $\mathbf{Img}_{\mathrm{tar}}$ and the retrieved
reference example.

Finally, the new segmentation result is stored back into memory, expanding the
knowledge base for future tasks:
\begin{equation}
\mathcal{M}_{t+1} = \mathcal{M}_{t} \cup \{(\mathbf{Img}_{\mathrm{tar}},\,
\mathbf{Mask}_{\mathrm{tar}})\},
\end{equation}
where $\mathcal{M}_{t}$ denotes the memory state after $t$ runs and
$\mathcal{M}_{t+1}$ represents the updated memory with the latest segmentation
added.

By leveraging task-relevant examples from memory, the system extends its capability to segment new instances of \textit{Golgi} automatically. Each new result enriches the memory, strengthening the knowledge base and improving future performance. As memory grows, subsequent tasks receive increasingly precise guidance, creating a virtuous cycle of experience accumulation, refinement, and evaluation. Over time, this leads to optimized workflow strategies, sharper evaluation prompts, and more effective tool selection, mirroring the iterative “experience acquisition → refinement → evaluation → decision-making” loop that formalizes the iterative paradigm driving self-evolving and continual learning LLMs. 


\subsection{Personalized System}\label{sec:personalize}

Our framework supports personalization, enabling users to balance the level of human-in-the-loop (HITL) interaction against inference time and result quality. At the end of each run, the system infers the user-preferred HITL level $\ell_{\mathrm{HITL}}$ from three modes, fully automatic, reference-guided, or human interactive based on the current run’s behavior and historical interaction patterns:
\begin{equation}
\ell_{\mathrm{HITL}}
= \mathcal{A}_{\mathrm{plan}}\!\left(\mathbf{s}_{\mathrm{current}},\, \mathbf{s}_{\mathrm{historical}}\right).
\end{equation}
Here, $\mathbf{s}_{\mathrm{current}}$ encodes user behavior during the present run (\eg tool usage, reference image/mask reliance, intervention frequency), while $\mathbf{s}_{\mathrm{historical}}$ aggregates long-term preferences across previous runs (\eg consistent trends in automation). See personalization prompts in Appendix~\ref{sec:appB:sum_personprompt}.

\section*{Data and code availability}\label{sec:avai}

\textbf{Datasets Description \& Repository}

The repository also contains a diverse set of datasets for both cell/nuclei and organelle segmentation.  

For \textbf{cells and nuclei}, the following datasets are included: \textbf{LiveCell}, which provides live-cell images with annotated cells and nuclei (\url{https://sartorius-research.github.io/LIVECell/}); \textbf{TissueNet}, a histological tissue dataset labeled for cells and nuclei (\url{https://deepcell.readthedocs.io/en/master/data-gallery/tissuenet.html}); \textbf{PlantSeg}, containing plant tissue images with segmentation labels (\url{https://github.com/tqwei05/PlantSeg}); \textbf{Lizard}, a dataset curated for cell and nuclei segmentation tasks (\url{https://www.kaggle.com/datasets/aadimator/lizard-dataset}), \textbf{2018 Data Science Bowl}~(\url{https://github.com/kamalkraj/DATA-SCIENCE-BOWL-2018}), a widely used nuclei segmentation benchmark from the 2018 Kaggle challenge, containing divergent microscopy images for automated nucleus detection; 
\textbf{Mouse Brain}~(\url{https://zenodo.org/records/11095111}), a dataset of cleared whole mouse brain nuclei imaged with a mesoSPIM system; and 
\textbf{Damond}~(\url{https://bodenmillergroup.github.io/imcdatasets/reference/Damond_2019_Pancreas.html}), an imaging mass cytometry dataset of human pancreas sections that includes single-cell data, multichannel images, and cell segmentation masks.

For \textbf{organelles}, the following datasets are included: the \textbf{CellMap Segmentation Challenge}, a benchmark dataset containing segmentation tasks for mitochondria, endoplasmic reticulum, and Golgi apparatus (\url{https://cellmapchallenge.janelia.org/}); \textbf{CEM-MitoLab}, a labeled collection of approximately 22,000 2D EM images with over 135,000 mitochondrial instances, extensively used for MitoNet training (\url{https://www.ebi.ac.uk/empiar/EMPIAR-11037/}); and the \textbf{ERNet dataset}, which provides endoplasmic reticulum images used for training and validation of ERNet models (\url{https://figshare.com/articles/dataset/ERnet_datasets}). 

\vspace{.2in}
\noindent \textbf{Code Availability}

\noindent The code and GUI are available at \url{https://github.com/yuxi120407/GenCellAgent}.

\section*{Acknowledgment}
This work was supported by the Laboratory Directed Research and Development (LDRD) Program 25-006 of Brookhaven National Laboratory under U.S. Department of Energy Contract No. DE-SC0012704. This research used resources of the National Synchrotron Light Source II, a U.S. Department of Energy (DOE) Office of Science User Facility operated for the DOE Office of Science by Brookhaven National Laboratory under Contract No. DE-SC0012704.

\section*{Author Contributions}
X. Y. designed and implemented the framework. Y. L. designed the framework and supervised the work. Y. Y., Q. L., Y. D., and S. M. provided the use case and discussed the framework.


\clearpage
\begin{appendices}
\section{Tools Repository}\label{sec:appA}


The repository integrates multiple categories of tools. The primary \textbf{LLM} used is \texttt{Gemini-2.5-Flash}, while the evaluation agent employs \texttt{Gemini-3.0-Flash-Preview} for processing both text and visual inputs. Both models are available at \url{https://deepmind.google/models/gemini/flash/}. For search functionalities, we utilize \textbf{Google Search} through the \textbf{SerpApi} interface (\url{https://serpapi.com}), which enables programmatic web queries and automated retrieval of relevant information.

Several \textbf{task-specific segmentation tools} are included. \textbf{MitoNet} is designed for mitochondrion segmentation (\url{https://volume-em.github.io/empanada.html}). \textbf{ERNet} focuses on endoplasmic reticulum segmentation (\url{https://github.com/charlesnchr/ERNet}). The \textbf{$\mu$SAM} framework provides segmentation for cells and nuclei (\url{https://github.com/computational-cell-analytics/micro-sam}), while \textbf{Cellpose} is another widely used generalist model for cell and nuclei segmentation (\url{https://github.com/MouseLand/cellpose}).

In addition, \textbf{general-purpose segmentation tools} are included. \textbf{SAM3} (Segment Anything with Concepts) can be accessed at \url{https://github.com/facebookresearch/sam3}. \textbf{SegGPT} (Segmenting Everything in Context) is available at \url{https://github.com/baaivision/Painter/tree/main/SegGPT}.

\section{2D Image Generation in CellMap Dataset}\label{sec:imggen}

The official CellMap segmentation dataset contains 289 annotated 3D crops in the training set, including 254 training crops and 35 validation crops, spanning multiple voxel resolutions from 4\,nm to 32\,nm. We use different subsets and 2D construction protocols depending on the evaluation objective. To assess the evaluation agent under diverse image conditions, we build the benchmark from the training split, which contains substantially greater diversity than the validation split. In contrast, for comparison with supervised baselines such as UNet, we use the official validation split to maintain a standard train/validation separation.

\textit{Training-split 2D construction for diversity evaluation:} For each 3D crop, we select the $z$-slice with the largest number of positive label pixels, discard slices without foreground pixels, crop to the valid non-NaN label region, remove low-contrast samples, and reduce redundancy using duplicate removal and SSIM-based filtering. Each retained slice is independently normalized to $[0,255]$ and saved together with its binarized mask. This procedure yields 103 mitochondria samples at $512 \times 512$ and 16$nm$, 85 ER samples at $256 \times 256$ and 8$nm$, and 35 Golgi samples at $512 \times 512$ and 4$nm$.

\textit{Validation-split 2D construction for supervised baseline comparison:} We construct the evaluation data directly from 2D validation patches generated by the CellMap dataloader. Each patch has size $128 \times 128$. Samples with fewer than 10 foreground pixels are discarded, each image is independently normalized to $[0,255]$, and redundant samples are removed using SSIM-based filtering with a threshold of 0.85. Black border regions are cropped before saving the final image--mask pairs. Using this procedure, we obtain 278 mitochondria samples at 8$nm$, 200 ER samples at 8$nm$, and 200, 110, and 64 Golgi samples at 8$nm$, 16$nm$ and 32$nm$, respectively.
\section{Details for Image Style Similarity Metric}\label{sec:style}

To quantify the similarity of visual style between two images, we adopt the approach introduced by~\cite{gatys2016image}, where style is represented using feature correlations extracted from a convolutional neural network (CNN). Specifically, the feature activations from layer $l$ of a pre-trained CNN are denoted as
\begin{equation}
F^{l} \in \mathbb{R}^{N_l \times M_l},
\label{eq:feature_activations}
\end{equation}
where $N_l$ is the number of feature maps in layer $l$ and $M_l$ is the number of spatial locations (\ie the product of height and width of the feature map). Each row of $F^{l}$ corresponds to a vectorized feature map.

The style representation of an image at layer $l$ is captured by the Gram matrix
\begin{equation}
G^{l} = F^{l} (F^{l})^{T} \in \mathbb{R}^{N_l \times N_l},
\label{eq:gram_matrix}
\end{equation}
where the $(i,j)$-th entry encodes the inner product between feature maps $i$ and $j$:
\begin{equation}
G^{l}_{ij} = \sum_{k=1}^{M_l} F^{l}_{ik} F^{l}_{jk}.
\label{eq:gram_entries}
\end{equation}
This captures correlations between different feature channels, which are indicative of the image’s texture and style.

To compare the style between two images $A$ and $B$, we compute their Gram matrices $G_A^l$ and $G_B^l$ at selected layers. Instead of using the mean squared difference (as in the original style transfer formulation), we normalize the matrices and compute the correlation coefficient between their entries. Let $\mathrm{vec}(G_A^l)$ and $\mathrm{vec}(G_B^l)$ denote the vectorized Gram matrices. The Pearson correlation coefficient is defined as
\begin{equation}
\rho^l(A,B) = \frac{\mathrm{cov}\!\left(\mathrm{vec}(G_A^l), \mathrm{vec}(G_B^l)\right)}{\sigma\!\left(\mathrm{vec}(G_A^l)\right) \, \sigma\!\left(\mathrm{vec}(G_B^l)\right)},
\label{eq:pearson_corr}
\end{equation}
where $\mathrm{cov}(\cdot,\cdot)$ is the covariance and $\sigma(\cdot)$ denotes the standard deviation.

The overall style similarity between the two images is then computed by averaging across the selected layers $\mathcal{L}$:
\begin{equation}
\mathrm{sim}(A,B) = \frac{1}{|\mathcal{L}|} \sum_{l \in \mathcal{L}} \rho^l(A,B).
\label{eq:overall_style_similarity}
\end{equation}

In our experiments, we use a pretrained VGG-19~\cite{simonyan2014very} network on ImageNet~\cite{krizhevsky2012imagenet} for style feature extraction. The style representation is computed using the following convolutional layers:
\[
\mathcal{L} = \{\text{conv1\_1}, \text{conv2\_1}, \text{conv3\_1}, \text{conv4\_1}, \text{conv5\_1} \}.
\]
These correspond to the first convolutional layer in each block of VGG-19. The final similarity score (Eq.~\eqref{eq:overall_style_similarity}) is the average Pearson correlation across these layers.

\section{Computational Cost Analysis}\label{sec:cost}

\begin{figure}[H]
 \centering 
\includegraphics[width=0.9
\textwidth]{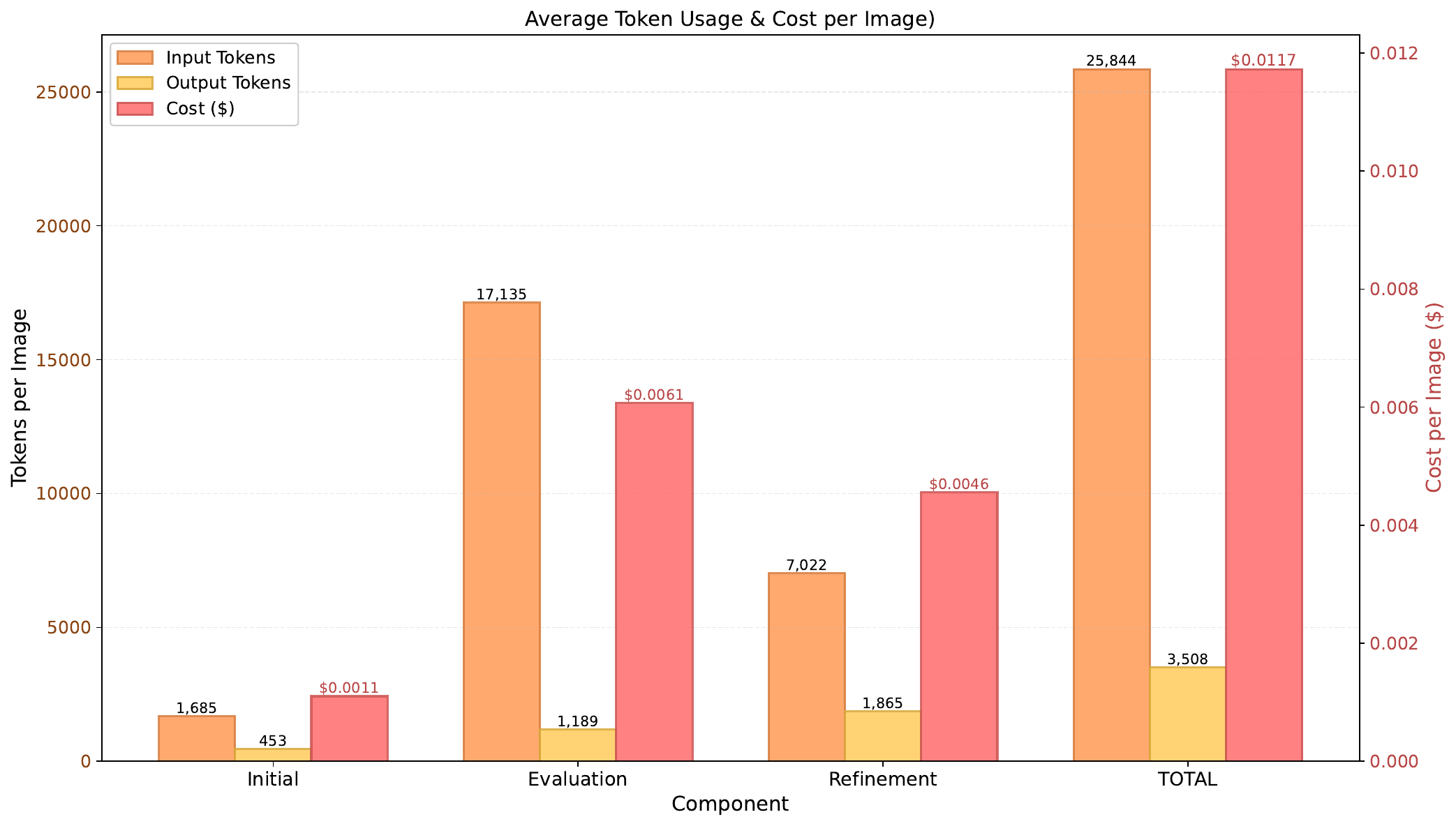}
\caption{ Average token usage and cost per image for different stages of the GenCELLAgent pipeline. Bars show input tokens, output tokens, and estimated API cost across the Initial prompting, Evaluation, and Refinement stages.}
 \label{fig:cost_summary}
\end{figure}

We now explicitly report token usage and the associated cost of the pipeline. For the text-guided general segmentation workflow, the average token usage per image is 25,844 input tokens and 3,508 output tokens across all stages (Fig.~\ref{fig:cost_summary}). Under the Gemini API pricing used in our experiments, this corresponds to an average total cost of approximately \$0.0117 per image, with the evaluation stage contributing the largest share (\$0.0061 per image). 



To assess the practical feasibility of GenCellAgent more broadly, we also analyze runtime and GPU usage of the built-in segmentation tools per image of $128\times128$. As shown in the Table~\ref{tab:runtime_memory}, their per-image runtime and GPU memory usage are as follows: MicroSAM: approximately 2.0 s/image with 4468 MB GPU memory; Cellpose: approximately 0.37 s/image with 1209 MB GPU memory; and CellSAM: approximately 0.64 s/image with 1476 MB GPU memory. These results show that the specialist-model path is substantially faster and lighter-weight than the iterative VLM-guided automatic mode, making it more suitable for routine in-domain segmentation, while the automatic mode is intended for more challenging out-of-domain or novel-object cases where specialist tools are unavailable or unreliable.

\begin{table}[t]
\centering
\caption{Runtime and GPU memory usage of segmentation components and built-in tools. The automatic mode consists of Gemini VLM bounding-box generation, SAM3 segmentation, and Gemini VLM evaluation/feedback.}
\label{tab:runtime_memory}
\begin{tabular}{lcc}
\hline
\textbf{Tool / Component} & \textbf{Real time per image} & \textbf{GPU Memory} \\
\hline
Gemini VLM (BBox generation) & 17.93 s & -- \\
SAM3 (Segmentation)& 2.82 s & 7818 MB\\
Gemini VLM (Evaluation + feedback) & 14.16 s & -- \\
\hline
Automatic mode total (per loop) & 34.91 s & -- \\
\hline 
Human Correction & 49.00 s & -- \\
Human-from-scratch Annotation & 227.00 s & -- \\
\hline
MicroSAM & 2.0 s & 4468 MB \\
Cellpose & 0.37 s & 1209 MB \\
CellSAM & 0.64 s & 1476 MB \\
\hline
\end{tabular}
\end{table}

\section{More Details for Human Interactions}\label{sec:appendix:HITL}

\begin{figure}[H]
 \centering 
 \includegraphics[width=\textwidth]{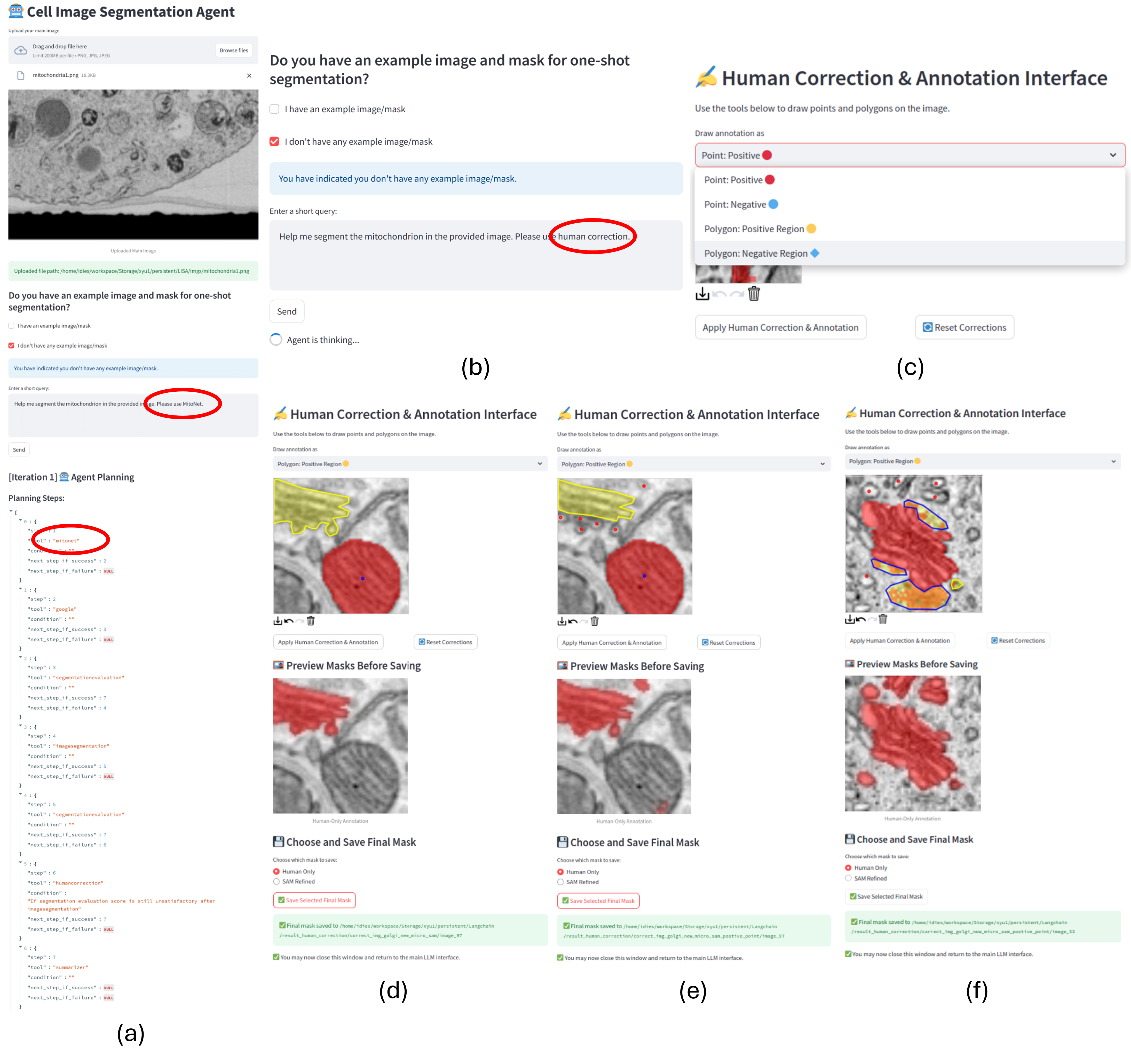}
 \caption{\textbf{Human-in-the-loop (HITL) framework in GenCellAgent}, zoomed in on an area of interest. (a) illustrating user query for a specific tool, initiation of the correction phase (b–c), and use of interactive point and polygon prompts for refining segmentation results (d–f).}
 \label{fig:HITL}
\end{figure}

The HITL process is illustrated  in Fig~\ref{fig:HITL}. The initial user interaction is shown in Fig.\ref{fig:HITL}a, where an expert provides a natural language query to GenCellAgent via the GUI. This query can specify the segmentation task and a preferred segmentation model. For example, the query, ``Help me segment the mitochondrion in the provided image. Please use MitoNet.'' directs the system to utilize a specific model, MitoNet, as highlighted by a red oval. GenCellAgent's planning agent processes this input and, if the request is reasonable, prioritizes the user's preference in its plan. The first step of the plan is thus to execute the MitoNet model for mitochondrial segmentation.

Users can initiate a correction phase by explicitly requesting human intervention, which prompts GenCellAgent to transition to a specialized manual annotation GUI, depicted in Fig.\ref{fig:HITL}b. This interface is equipped with a suite of tools for correcting segmentation errors and annotating new objects. The core tools include point- and polygon-based prompts, which can be either positive or negative. \textbf{Positive Point Prompts}: A positive point prompt allows users to add regions to an existing segmentation, expanding the object boundary where needed. \textbf{Negative Point Prompts}: A negative point prompt provides a quick and efficient way to remove an entire segmented object that contains the selected point with a single click. \textbf{Positive Polygon Prompts}: For complex or challenging regions, a positive polygon allows the user to delineate a new object, which is then segmented by the $\mu$SAM model. \textbf{Negative Polygon Prompts}: A negative polygon prompt enables the user to precisely remove a selected portion of a segmented object without deleting the entire instance. This is particularly useful for fine-tuning object boundaries or eliminating false-positive regions.

The effectiveness of these tools is demonstrated in Fig.\ref{fig:HITL}d-f. Fig.\ref{fig:HITL}d shows how a user can add a new object using a positive polygon (yellow) and remove an entire, unwanted object with a single click using a negative point (blue). Fig.\ref{fig:HITL}e illustrates the efficiency of adding multiple new objects quickly, using a simple positive polygon and a couple of positive points (red), leveraging the speed and accuracy of $\mu$SAM. Fig.\ref{fig:HITL}f highlights the precision of the polygon tools, showing how a negative polygon (blue) can be used to remove false-positive regions from a segmented object, while positive points (red) add other objects. This HITL framework guarantees that GenCellAgent can be easily and effectively guided by human expertise, leading to superior and more trustworthy segmentation results for complex biological images.


\section{GUI Examples for Fully Automatic Process}

\begin{figure}
 \centering 
 \includegraphics[width=\textwidth]{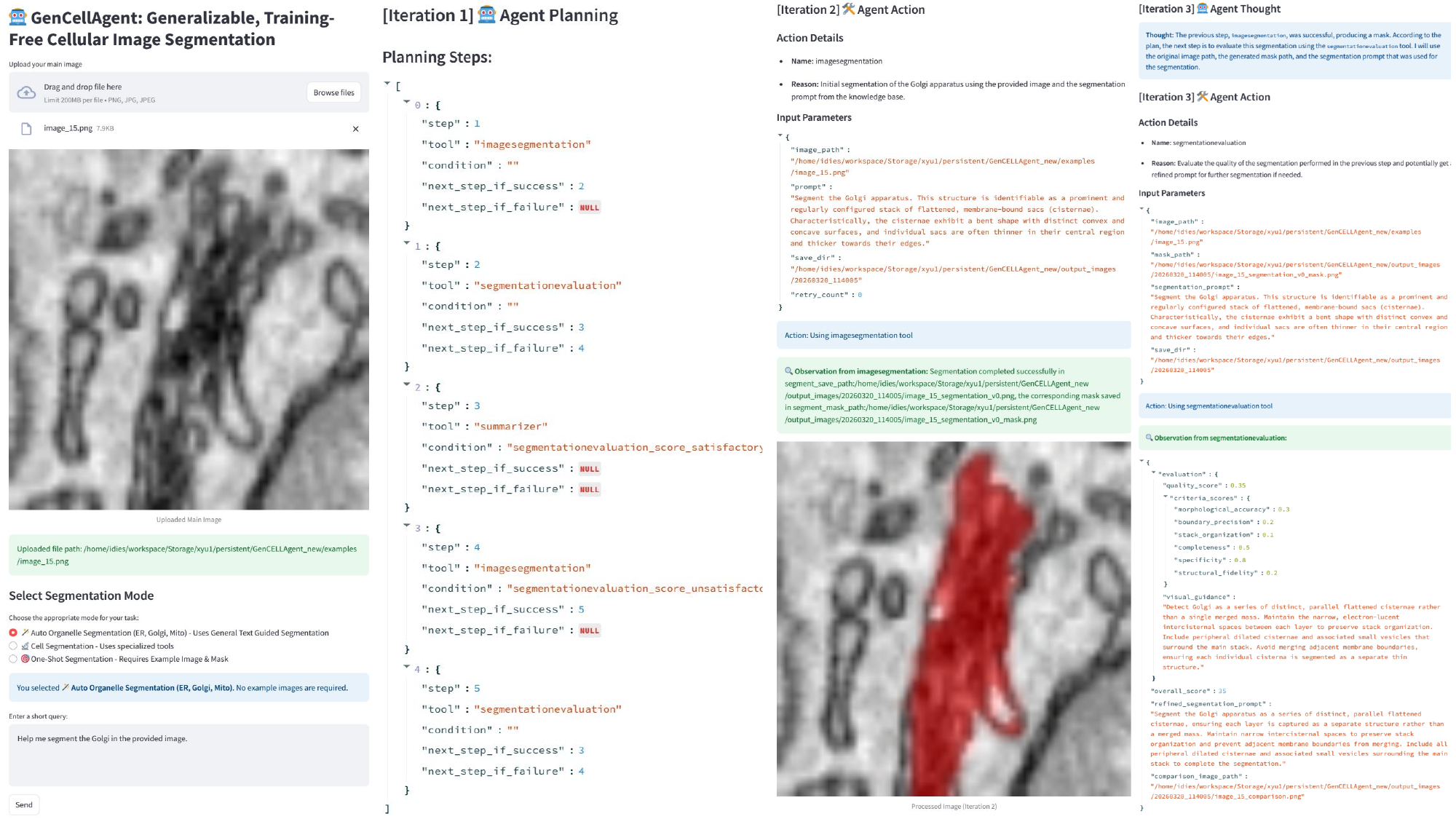}
 \caption{\textbf{GenCellAgent GUI example at the start of a Golgi image segmentation workflow}, zoomed in on an area of interest, showing image upload, agent planning, retrieval of Golgi visual characteristics in memory, and the initial results.}
 \label{GUI_screens_1}
\end{figure}

\begin{figure}
 \centering 
 \includegraphics[width=\textwidth]{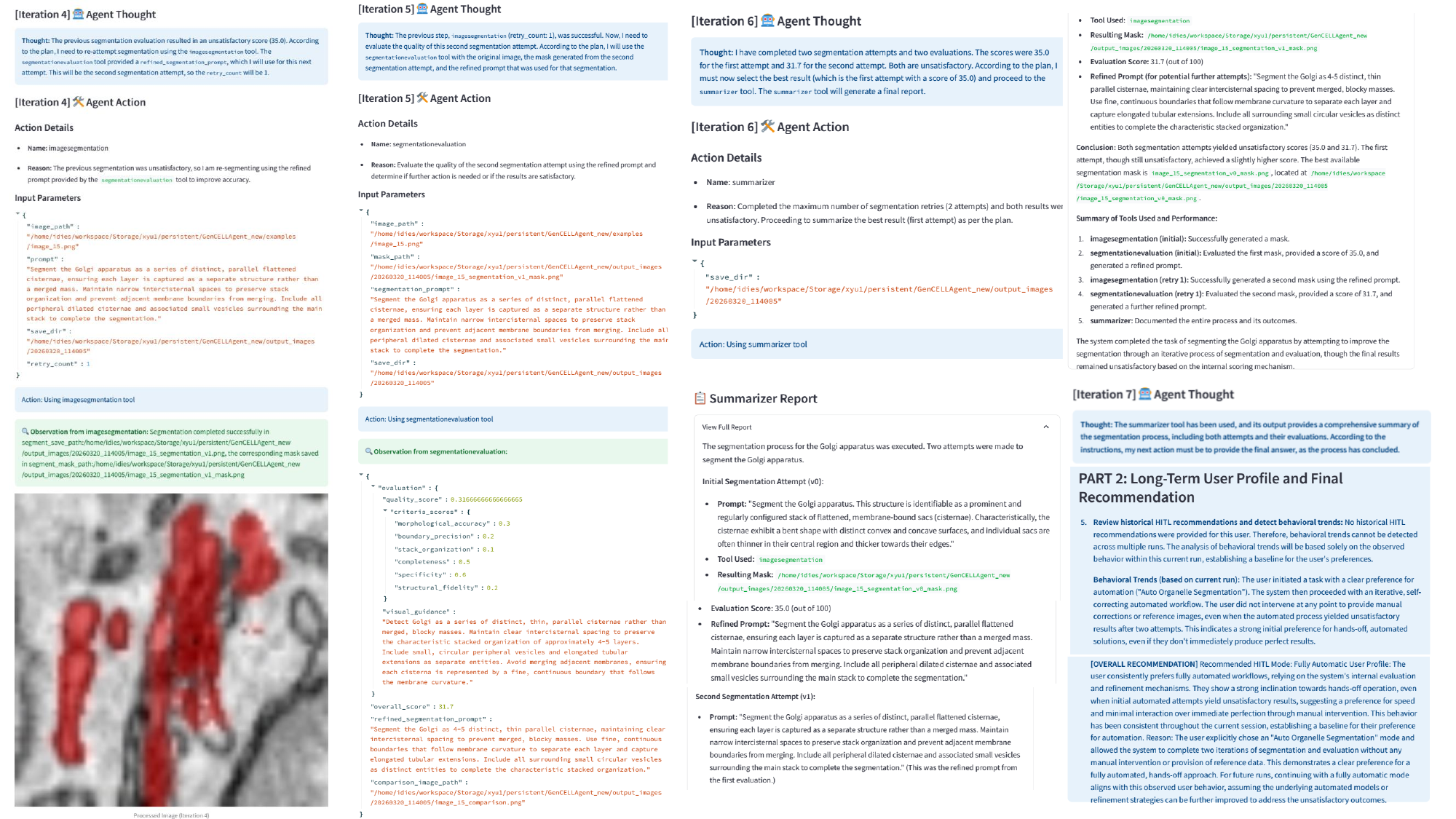}
 \caption{\textbf{GenCellAgent GUI example for illustrating the automated loop for Golgi image segmentation}, zoomed in on an area of interest, including evaluation of initial results (scoring and prompt refinement), feedback-driven tool re-runs, iterative re-evaluation, summary report and personalized recommendations.}
 \label{GUI_screens_2}
\end{figure}

\section{Prompts}\label{sec:prompt}


\subsection{Execution Prompt}\label{sec:appB:exeprompt}

\begin{ExecPrompt}
You are a ReAct (Reasoning and Acting) agent tasked with answering the following query:

Query: {query}

The planning: {planning}

The historical_information: {historical_information}

Your goal is to reason about the query and decide on the best course of action to answer it accurately.

Previous reasoning steps and observations: {history}

Current Segmentation Retry Count: {retry_count} / {max_retries}

Available tools: {tools}

Instructions:
1. Analyze the query, previous reasoning steps, and observations.
2. Decide on the next action: use a tool or provide a final answer.
3. Respond in the following JSON format. Your entire response MUST be a single, valid JSON block. Do not add conversational text like "Thought:" or "Action:" before the JSON.

If you need to use a tool:
{{
    "thought": "Your detailed reasoning about what to do next",
    "action": {{
        "name": "Tool name (google, imagesegmentation, oneshotsegmentation, segmentationevaluation)",
        "reason": "Explanation of why you chose this tool",
        "input": "Specific input for the tool, if different from the original query"
    }}
}}

For google tool, this is a web search tool to search the visual characteristics—such as shape, texture, and location—rather than functionality of the segment object, please using following formate:

{{
  "thought": "Your detailed reasoning about what to do next",
  "action": {{
    "name": "google",
    "reason": "Explanation of why you chose this tool",
    "input": {{
      "search_query": "<insert the search of visual characteristics—such as shape, texture, and location—rather than functionality of the segment object in the query>",
      "location":''
    }}
  }}
}}


For oneshotsegmentation tool, this is a few-shot image segmentation tool (SegGPT) used when the user provides a reference image and mask. Please use the following JSON format:

{{
  "thought": "Your detailed reasoning about why one-shot segmentation is appropriate",
  "action": {{
    "name": "oneshotsegmentation",
    "reason": "Explanation of why you chose this tool",
    "input": {{
      "image_path": "<insert the image_path provided in the query>",
      "prompt_image_path": "<insert the prompt_image_path provided in the query>",
      "prompt_mask_path": "<insert the prompt_mask_path provided in the query>",
      "save_dir": "<insert the save_directory provided in the query>"
    }}
  }}
}}


For imagesegmentation tool, this is a general image segmentation tool that uses Gemini to generate bounding boxes and SAM3 to segment. Please use the following JSON format:

{{
  "thought": "Your detailed reasoning about the description of the segment object",
  "action": {{
    "name": "imagesegmentation",
    "reason": "Explanation of why you chose this tool",
    "input": {{
      "image_path": "<insert the image_path provided in the query>",
      "prompt": "<insert the segmentation_prompt obtained from the search tool's output when first use or insert the refined_segmentation_prompt from the segmentationevaluation tool's output when second use>",
      "save_dir": "<insert the save_directory provided in the query>",
      "retry_count": {retry_count}
    }}
  }}
}}



For segmentationevaluation tool, this is a segmentation performance evaluation tool using Gemini VLM. It evaluates the mask quality and returns a score plus a refined segmentation prompt. Please use the following format:

{{
  "thought": "Your detailed reasoning about the evaluation of the segment object image",
  "action": {{
    "name": "segmentationevaluation",
    "reason": "Explanation of why you chose this tool",
    "input": {{
      "image_path": "<insert the original image_path provided in the query>",
      "mask_path": "<insert the segment_mask_path from the imagesegmentation tool's output>",
      "segmentation_prompt": "<insert the segmentation_prompt obtained from the search tool's output, OR the original user query if no search was performed, OR the refined_segmentation_prompt from the previous segmentationevaluation tool's output>",
      "save_dir": "<insert the save_directory provided in the query>"
    }}
  }}
}}


For summarizer tool, this is the summarize tool only used in the final to generate the report, please using following format:

{{
  "thought": "Your detailed reasoning about what to do next",
  "action": {{
    "name": "summarizer",
    "reason": "Explanation of why you chose this tool",
    "input": {{
      "save_dir": "<insert the save_directory provided in the query>"
    }}
  }}
}}




If you have enough information to answer the query:
{{
    "thought": "Your final reasoning process",
    "answer": "Your comprehensive answer to the query"
}}

Remember:
- Be thorough in your reasoning.
- Use tools when you need more information.
- Always base your reasoning on the actual observations from tool use.
- If a tool returns no results or fails, acknowledge this and consider using a different tool or approach.
- Provide a final answer only when you're confident you have sufficient information.
- If you cannot find the necessary information after using available tools, admit that you don't have enough information to answer the query confidently.
- Finally, use the summarizer tool to generate the summarized report.
- Always follow segmentation → evaluation → segmentation → evaluation strictly.
- Avoid repeating the same tool consecutively (every segmentation must be followed by evaluation, and every evaluation must be followed by segmentation unless finalization).

1. Start with the Google tool: Use the google search tool to:
  -- Retrieve visual characteristics of the segmented object (e.g., shape, texture, structure, location).
  -- Generate the initial segmentation prompt, which will be passed to the planned segmentation tool.
  -- IMPORTANT: If the target organelle's visual characteristics and segmentation prompt are already present in your `historical_information` (specifically within the "KNOWLEDGE BASE" section), DO NOT use the google tool. Skip this step and immediately use the appropriate segmentation tool (as defined by your plan and the user's selected mode) using the saved `segmentation_prompt` from memory.

2. Run the chosen segmentation tool:
  -- Use the segmentation prompt obtained from the Google tool as input.
  -- Produce an initial segmentation result.


3. Evaluate the result using the segmentationevaluation tool:
  -- Evaluation is mandatory after every segmentation.
  -- Assess the quality of the segmentation.
  -- Generate a refined segmentation prompt based on the evaluation.

4. Decision logic based on evaluation:
  -- If performance is satisfactory → proceed directly to the summarizer tool and terminate.
  -- If performance is unsatisfactory:
      -- Strictly re-enter the segmentation → evaluation loop.
      -- Use the refined segmentation prompt from segmentationevaluation as new input to the imagesegmentation tool.
      -- Re-run segmentation and reevaluate.
      
5. Repeat the imagesegmentation and segmentationevaluation cycle:
  -- Segmentation must always be followed by evaluation.
  -- Always use the most recent refined prompt to inform the next segmentation.
  -- Stop only when the evaluation tool indicates satisfactory performance OR if you have completed EXACTLY {max_retries} segmentation attempts.
  -- MAXIMUM RETRY LIMIT: You must run the `imagesegmentation` tool up to {max_retries} times if the score remains unsatisfactory. Do not stop early just because a score decreases; use all {max_retries} attempts to try and find a better prompt. 
  -- Once you evaluate the final {max_retries}th attempt, if the score is still unsatisfactory, DO NOT try to segment again. Instead, look at the scores from all {max_retries} attempts, select the best result, and immediately proceed to the `summarizer` tool.
  
6. Finalization:
  -- Once segmentation is deemed satisfactory, OR you hit the {max_retries}-attempt limit, you MUST invoke the summarizer tool to produce a final summary of the process and results.
  -- IMPORTANT: You may only call the `summarizer` tool EXACTLY ONCE. 
  -- If you see that the `summarizer` tool has already been used and its observation is in the history, your VERY NEXT ACTION MUST be to output your final `"answer"` to end the interaction. Do not use any more tools.

\end{ExecPrompt}
\captionof{listing}{Execution Prompt}

\subsection{Planning Prompt}\label{sec:appB:planprompt}
\begin{PlanningPrompt}
You are a planning agent responsible for constructing a fully automated segmentation workflow to address the following query:

Query: {query}

Previously saved object information: {saved_information}

You have access to the following tools: {tools}

[{{
    "name": "google",
    "description": "This is a web search tool to search the text description of visual characteristics—such as shape, texture, and location—rather than functionality of the segment object when Previously saved object information is None. Always use it once before the first using segmentationevaluation."
  }}

  {{
    "name": "imagesegmentation",
    "description": "This is the segmentation tool. General-purpose segmentation tool used when no specific tool is available or the specific tool doesn't work well."
  }},

  {{
    "name": "oneshotsegmentation",
    "description": "Few-shot image segmentation tool (SegGPT). Use this directly if the user has provided a reference prompt image and prompt mask."
  }},

  
  {{
    "name": "segmentationevaluation",
    "description": "This is the segmentation evaluation tool, evaluate the performance of a segmentation result based on visual characteristics."
  }},
  

  {{
    "name": "summarizer",
    "description": "Generates a final report summarizing the entire segmentation process."
  }}
]


Your task:
- Plan a fully autonomous segmentation workflow by defining the high-level sequence of tools. Do NOT formulate the specific inputs for the tools in this plan; only define the steps, conditions, and tool names.
- If reference images (`prompt_image_path` and `prompt_mask_path`) are provided in the query, prioritize using the `oneshotsegmentation` tool.
- If saved information (including the KNOWLEDGE BASE) includes matching object name and valid visual characteristics or reference data, plan to use the appropriate segmentation tool directly and SKIP the google search.
- If not, and no reference is provided, plan to retrieve characteristics using the google tool (used once only), and proceed accordingly.
- Evaluate segmentation quality using segmentationevaluation and adjust strategy based on score.
- Apply fallback segmentation tools only if the evaluation score is unsatisfactory.
- Always end with the summarizer tool to produce a detailed process report.

**Important Rules**:
- No human interaction is allowed during this process.
- Use conditional logic based on segmentationevaluation scores to guide tool usage.
- Avoid redundancy and always end with a summarizer.

1. Initial Path Selection:
  -- If reference images are provided: Start the plan with `oneshotsegmentation`.
  -- If NO reference is provided AND the object's characteristics are ALREADY in the KNOWLEDGE BASE: Skip `google` and start the plan directly with the appropriate segmentation tool.
  -- If NO reference is provided AND the object is NOT in the KNOWLEDGE BASE: Start the plan with `google` search to gather visual characteristics.

2. Run Segmentation:
  -- Execute the chosen segmentation tool (`oneshotsegmentation` or `imagesegmentation`).
  -- Produce an initial segmentation result.


3. Evaluate the result using the segmentationevaluation tool:
  -- Evaluation is mandatory after every segmentation.
  -- Assess the quality of the segmentation.
  -- Generate a refined segmentation prompt based on the evaluation.

4. Decision logic based on evaluation:
  -- If performance is satisfactory → proceed directly to the summarizer tool and terminate.
  -- If performance is unsatisfactory:
      -- Strictly re-enter the segmentation → evaluation loop.
      -- Use the refined segmentation prompt from segmentationevaluation as new input to the imagesegmentation tool.
      -- Re-run segmentation and reevaluate.
      
5. Repeat the imagesegmentation and segmentationevaluation cycle:
  -- Segmentation must always be followed by evaluation.
  -- Always use the most recent refined prompt to inform the next segmentation.
  -- Stop only when the evaluation tool indicates satisfactory performance.
  
6. Finalization:
  -- Once segmentation is deemed satisfactory, invoke the summarizer tool to produce a final summary of the process and results.
  
Additional Notes
- Every iteration should build upon previous insights.
- Avoid repeating the same prompt or tool usage without updated input.
- Use tools only when necessary, driven by evaluation feedback.

Output Format (MUST follow this JSON structure):
[
  {{
    "step": <integer, step number>,
    "tool": "<tool_name>",
    "condition": "<optional condition for this step (leave blank if unconditional)>",
    "next_step_if_success": <next step number (null if final step)>,
    "next_step_if_failure": <alternative next step number if applicable>
  }},
  ...
]

\end{PlanningPrompt}
\captionof{listing}{Planning Prompt}

\subsection{Evaluation \& Feedback Prompt}\label{sec:appB:evaprompt}
\begin{EvaluationPrompt}
You are an expert in electron microscopy analyzing Golgi apparatus segmentation quality.

IMAGE LAYOUT:
- LEFT: Original electron microscopy image showing Golgi apparatus structures
- RIGHT: Current segmentation (WHITE = detected as Golgi, BLACK = background)

TASK: Evaluate segmentation quality and provide concise Golgi detection guidance.

STEP 1: EVALUATE QUALITY BASED ON SPECIFIC CRITERIA

Evaluate by comparing LEFT to RIGHT across six criteria (each 0.0-1.0):

1. MORPHOLOGICAL ACCURACY: Are all Golgi components (cisternal stacks, individual cisternae, associated vesicles) captured?
2. BOUNDARY PRECISION: Do boundaries tightly follow cisternal membrane edges without merging adjacent cisternae?
3. STACK ORGANIZATION: Are parallel cisternae maintained as separate layers with proper intercisternal spacing?
4. COMPLETENESS: What percentage of visible Golgi structures in LEFT (including peripheral regions and vesicles) is captured in RIGHT?
5. SPECIFICITY: Is ONLY Golgi segmented (no ER, mitochondria, endosomes, or other organelles)?
6. STRUCTURAL FIDELITY: Is the characteristic stacked, polarized architecture (cis-to-trans) preserved?

quality_score = average of 6 criteria

STEP 2: PROVIDE CONCISE GOLGI DETECTION GUIDANCE

Based on comparing LEFT to RIGHT, write 3-5 sentences describing what Golgi SHOULD look like for detection.
Focus on what the current segmentation needs to capture better.

OUTPUT FORMAT (JSON only, no markdown, no backticks):
{{
  "quality_score": <float 0.0-1.0>,
  "criteria_scores": {{
    "morphological_accuracy": <float 0.0-1.0>,
    "boundary_precision": <float 0.0-1.0>,
    "stack_organization": <float 0.0-1.0>,
    "completeness": <float 0.0-1.0>,
    "specificity": <float 0.0-1.0>,
    "structural_fidelity": <float 0.0-1.0>
  }},
  "visual_guidance": "3-5 sentences describing Golgi visual characteristics for detection, based on weaknesses in current segmentation.

Example format (adapt to this specific image):
'Golgi appears as [describe stacks and cisternae]. Detect [number] separate cisternae with [boundary characteristics]. Maintain [spacing between layers]. Include [peripheral structures like vesicles]. Avoid [incorrectly merged or misclassified structures].'

Be specific to THIS image's Golgi and current segmentation issues.
Use imperative detection language: 'Detect...', 'Include...', 'Maintain...', 'Avoid...'
NO coordinates or pixel locations."
}}

Keep it concise - focus on the most critical detection guidance based on current results.

\end{EvaluationPrompt}
\captionof{listing}{Evaluation Prompt}

\subsubsection{Refined Segmentation Prompt}\label{sec:appB:refsegprompt}
\begin{RefinePrompt}
Analyze this microscopy image to identify Golgi apparatus structures for SAM3 segmentation.

TASK: Provide detailed prompts for SAM3 (Segment Anything Model 3) to accurately segment all Golgi apparatus regions.

WHAT TO DETECT:
{feedback}

PROMPT REQUIREMENTS:
1. positive_boxes: Bounding boxes [ymin, xmin, ymax, xmax] around distinct Golgi regions
   - FIRST: Analyze the image and estimate how many distinct Golgi apparatus regions are visible
   - THEN: Provide one bounding box for each Golgi complex or distinct Golgi region
   - Cover all visible Golgi structures that are clearly distinguishable
   - Make boxes tight around the Golgi apparatus
   - Include the main Golgi stack and closely associated vesicles
   - AVOID including other organelles (mitochondria, ER, nuclei, large vesicles)
   - Focus on regions with clear Golgi morphology (stacked/layered appearance)
   - Each box should contain primarily Golgi apparatus, not mixed structures
   - Number of boxes should match the number of distinct Golgi regions you identify

2. positive_points: [y, x] coordinates INSIDE clear Golgi structures (6-12 points)
   - Place points in the center of Golgi stacks
   - Focus on the densest/darkest parts of the Golgi apparatus
   - Distribute across different Golgi cisternae if multiple stacks are visible
   - Include points in characteristic crescent or curved regions
   - MORE positive points = better segmentation

3. negative_points: [y, x] coordinates in NON-Golgi regions (12-20 points)
   - Place in nuclei (large dark circular regions)
   - Place in ER (reticular network structures)
   - Place in mitochondria (small oval organelles)
   - Place in large vesicles and vacuoles
   - Place in background/cytoplasm (areas without Golgi)
   - Place in regions that might be confused with Golgi (dense ER regions)
   - Place near Golgi boundaries to refine edges
   - MORE negative points = cleaner segmentation, fewer false positives

COORDINATE FORMAT:
- All coordinates normalized to [0, 1000] range
- Based on image dimensions: [0, 1000] maps to [0, width] and [0, height]
- Example: center of image = [500, 500]

OUTPUT FORMAT:
Return ONLY valid JSON (no markdown, no backticks, no extra text):
{{
  "positive_boxes": [[ymin, xmin, ymax, xmax], ...],
  "positive_points": [[y, x], ...],
  "negative_points": [[y, x], ...]
}}

CRITICAL: 
- First assess the image to determine the appropriate number of bounding boxes
- Look for characteristic stacked/layered morphology
- Provide appropriate points (6-12 positive, 12-20 negative) for accurate SAM3 segmentation
- Ensure boxes exclude other organelles to avoid contamination
- Adapt the number of boxes based on Golgi distribution and visibility"""
\end{RefinePrompt}
\captionof{listing}{Refined Segmentation Prompt}

\subsection{Summarize \& Personalization Prompt}\label{sec:appB:sum_personprompt}
\begin{SummarizePrompt}
You are an intelligent assistant summarizing a segmentation workflow to understand user behavior and recommend the appropriate HITL mode for the next run. Your output must be based on **both the current session and historical HITL recommendations**, with a strong focus on user behavior progression.

## Current Run
- Query: {query}
- Interaction History:
{history}

## Historical HITL Mode Recommendations:
The following recommendations were made in previous runs:
{previous_hitl_results}

---

## PART 1: Current Run Analysis
1. Describe the segmentation process used in the current run.
2. List the tools used in order and what each contributed.
3. Summarize the user’s interaction behavior:
   - Use of automatic vs manual tools
   - Use of references (e.g. one-shot segmentation)
   - Feedback frequency
   - Number of iterations

4. Based on this run alone, recommend:
[CURRENT RUN]
Recommended HITL Mode: <Fully Automatic | Reference Guided | Human Interaction>
Reason: <why this HITL mode fits this specific run>

---

## PART 2: Long-Term User Profile and Final Recommendation
5. Review the historical HITL recommendations and detect **behavioral trends**:
- Is the user becoming more or less interactive over time?
- Are they consistently using the same tools or exploring new ones?
- Are they gradually shifting from automation to correction (or vice versa)?

6. Generate a long-term **User Profile** considering both the current and past sessions.
Example profiles:
- "Consistently prefers fully automated workflows with minimal feedback."
- "Has evolved from reference-based guidance to more manual correction."
- "Initially used correction tools but now prefers faster automatic approaches."

7. Provide the final recommendation:

[OVERALL RECOMMENDATION]
Recommended HITL Mode: <Fully Automatic | Reference Guided | Human Interaction>
User Profile: <summary across runs that includes progression or consistency>
Reason: <why this mode is appropriate based on the pattern across sessions>
--

## Guidance:
- If the tool `oneshotsegmentation` was used in the current run, the user clearly provided a reference image and mask.  
  → In that case, recommend `Reference Guided` for [CURRENT RUN].
- Use other tool choices (like `humancorrection`, `mitonet`, `imagesegmentation`, etc.) and the historical HITL trends to assess consistency or change.
- Do not recommend a different HITL mode unless there is clear evidence the user's behavior has shifted.
- In the `[OVERALL RECOMMENDATION]`, the user profile must reflect their **behavior evolution** over time.
- Include progression insight in the User Profile: e.g., "The user increasingly engages with manual tools."


The final output must include:
- A `[CURRENT RUN]` block  
- A `[OVERALL RECOMMENDATION]` block  
- A thoughtful User Profile that reflects both behavior **and change over time**

\end{SummarizePrompt}
\captionof{listing}{Summarize \& Personalization Prompt}

\subsubsection{Search Summarize Prompt}\label{sec:appB:searchprompt}
\begin{SearchPrompt}
Analyze the following search results and do two things:
1. Summarize the Visual Characteristics described across the search content.
2. Generate a Segmentation Prompt that could be used to guide a visual segmentation tool
based on those characteristics.
{search content}
Output format:\n"
### Visual Characteristics Summary ###
[your summary here]
### Segmentation Prompt ###
[your segmentation prompt here]
\end{SearchPrompt}
\captionof{listing}{Search Summarize Prompt}
\end{appendices}

\end{document}